\documentclass[12pt]{article}
\usepackage[utf8]{inputenc}

\usepackage{amsthm,amsmath}
\usepackage[authoryear,longnamesfirst]{natbib}
\usepackage[colorlinks,citecolor=blue,urlcolor=blue]{hyperref}
\usepackage{setspace}
\usepackage{authblk}
\usepackage{amssymb}
\usepackage{graphicx}
\usepackage{amsfonts}
\usepackage{float}
\usepackage[normalem]{ulem}
\usepackage{multirow}
\usepackage{upgreek,textgreek}
\usepackage[para, flushleft]{threeparttable}
\usepackage[margin=1in]{geometry}

\allowdisplaybreaks

\newcommand{\bpsi}{ \mbox{\boldmath $\psi$}}

\newcommand{\btheta}{ \mbox{\boldmath $ \theta $} }

\newcommand{\bgamma}{ \mbox{\boldmath $\gamma$} }

\newcommand{\bone}{\textbf{1}}
\newcommand{\bZ}{\textbf{Z}}

\newcommand{\bs}{\textbf{s}}

\newcommand{\bV}{\textbf{V}}

\newcommand{\bY}{\textbf{Y}}

\title{Joint Spatial and Temporal Generalized Dissimilarity Mixed Modeling (stGDMM) for Beta Diversity}
\author[1]{Philip A. White\thanks{Corresponding author. Conceptualization, Methodology, Formal Analysis, Software, and Writing (original draft, review \& editing).}}
\author[2]{Henry A. Frye\thanks{Conceptualization and Writing (review \& editing).}}
\author[3,4]{Jasper A. Slingsby\thanks{Conceptualization, Data curation, and Writing (review \& editing).}}
\author[5]{John A. Silander, Jr.\thanks{Conceptualization and Writing (review \& editing).}}
\author[3,4]{Hana Petersen\thanks{Conceptualization, Data curation, and Writing (review \& editing).}}
\author[6]{Alan E. Gelfand\thanks{Conceptualization, Methodology, and Writing (original draft, review \& editing).}}

\affil[1]{Department of Statistics, Brigham Young University, Provo, UT, USA}
\affil[2]{Forest and Wildlife Ecology, University of Wisconsin-Madison, Madison, WI, USA}
\affil[3]{Department of Biological Sciences and Centre for Statistics in Ecology, Environment and Conservation, University of Cape Town, Cape Town, South Africa}
\affil[4]{Fynbos Node, South African Environmental Observation Network, Cape Town, South Africa}
\affil[5]{Department of Ecology \& Evolutionary Biology, University of Connecticut, Storrs, CT, USA}
\affil[6]{Department of Statistical Science, Duke University, Durham, NC, USA}

\begin{document}

\maketitle

\begin{abstract}
Generalized dissimilarity models (GDMs) have emerged as a valuable tool for formal statistical analysis of biodiversity.  In particular, beta diversity, measured using dissimilarity measures, e.g., the Bray-Curtis dissimilarity in our case, provides a statistical summary of the difference in species composition between sites.  It also provides novel data for spatial and spatio-temporal modeling as it resides over the product space of one space-time pair and a second space-time pair.  In earlier work we developed the spatial generalized dissimilarity mixed model (spGDMM) to remedy some of the stochastic issues concerned with the foundational GDM in the literature.  Here, we extend that work to include dynamics. We find much richer modeling opportunities as we consider beta diversity with regard to change in time as well as space.  We illustrate with a dataset from the Cape Floristic Region (CFR) in South Africa.
\end{abstract}


\section{Introduction}

There has long been a central focus of community ecology on predicting and understanding changes in species composition over space and time. Spatial variation in species composition of communities has been termed beta diversity, while alpha diversity references taxonomic variation at a single location \citep{whittaker_evolution_1972}. Beta diversity has become a key measure of spatial variation in biodiversity \citep{mcgill_fifteen_2015}. With the relatively recent recognition that human activity across the planet has increasingly affected broad natural patterns on biodiversity \citep{cardinale_biodiversity_2012, johnson_biodiversity_2017, jaureguiberry_direct_2022}, there has been a call for studies to better understand how biodiversity has changed across spatial scales and taxonomic groups, evaluate biodiversity metrics, and develop new tools to model biodiversity \citep{mcgill_fifteen_2015,dornelas_looking_2023,heino2024navigating}.


Although there has been considerable effort in modeling spatial patterns in beta diversity (See \cite{white2024generative} and reference therein), developing sufficient tools for modeling temporal patterns in beta diversity has been much more recalcitrant \citep{legendre2019temporal, magurran_temporal_2019, chao_measuring_2021}, and indeed there has been little effort for jointly modeling spatio-temporal patterns in beta diversity. Statistical modeling of beta diversity is challenging because the response is typically defined through pairwise dissimilarities, inducing dependence and requiring models that can accommodate nonlinear environmental relationships. Generalized Dissimilarity Modeling (GDM) was introduced as a flexible framework for relating compositional turnover to geographic and environmental differences \citep{ferrier2002mapping, ferrier2007using} and has become one of the main approaches that ecologists use to model beta diversity. However, the standard GDM does not provide a fully generative probabilistic model for pairwise dissimilarities, leaving several inferential and uncertainty-quantification challenges. To address these issues, \cite{white2024generative} developed a stochastic modeling framework for beta diversity that embeds GDM-style environmental warping within a formal hierarchical model.

With global environmental change there is increasing emphasis on biodiversity observation over time to detect signals of change from natural dynamics, infer potential drivers, and identify factors or interventions that may convey resistance or resilience in the face of changing environmental conditions \citep{scholes2008, diaz2015}. This imperative has promoted the development and accumulation of biodiversity time-series datasets \citep{dornelas2018, dornelas2025} and methods with which to analyze them \citep{legendre2019temporal}. 
\cite{heino2024navigating} distinguish four spatial and temporal perspectives on beta diversity that are relevant to ecological and biodiversity applications: spatial beta diversity, temporal beta diversity, temporal variation in spatial beta diversity, and spatial variation in temporal beta diversity. \cite{heino2024navigating} further note that each of these perspectives can be studied using presence--absence or abundance data, and can be defined with respect to taxonomic, phylogenetic, or functional community features. Thus, jointly accounting for spatial and temporal dynamics is essential to providing a more complete picture of beta diversity.

In this paper, our contributions are focused on addressing these components of beta diversity by extending the framework proposed by \cite{white2024generative} to include dynamics.  
Specifically, working with dissimilarity measures, we find ourselves having to model data which lies in a product space of (time, space) $\times$ (time, space). We outline the advantages of our approach in Section \ref{sec:contributions}. The data come from fifty \(5 \times 10\) m\(^2\) permanent relev\'es in the Cape Peninsula of South Africa, surveyed twice, once in 1996 and again in 2021. Treating each site-year combination as an observation yields 100 site-years and ${100 \choose 2} = 4950$ pairwise Bray--Curtis dissimilarities. The Cape peninsula is part of the Cape Floristic Region, a biodiversity hotspot \citep{Myers2000-nc} uniquely characterized by its exceptionally high compositional turnover \citep{latimer_neutral_2005}. Despite recent efforts to assimilate datasets to better understand the temporal components of beta diversity, e.g., \citep{dornelas2018}, ecologists are limited to what has been historically sampled. Given the logistical difficulties in sampling high biodiversity ecosystems, the number of sampled time points in these regions is often limited despite their out-sized importance to global biodiversity. Our dataset with two time points is illustrative of a scenario that ecologists often face where the availability of repeated sampling is limited to only a few time points, e.g., \citep{pauli_recent_2012, dalmaso_spatial_2020}.




This dataset motivates our methodological development in several ways. First, because the response is a pairwise dissimilarity among site-years, the data naturally live on the product space $(t,\bs)\times(t',\bs')$, inducing dependence among observations that share sites, years, or both. Second, the two surveys are separated by 25 years, creating a setting in which beta diversity may exhibit both temporal persistence and temporal change, requiring a model that can borrow information across years while still allowing ecological relationships to evolve. Third, the covariates include spatially varying environmental gradients, an ordinal moisture variable, a time-varying disturbance variable, and geographic distance, motivating flexible warped covariate effects that can accommodate nonlinear and potentially time-dependent relationships with compositional turnover. Finally, the presence of exact Bray--Curtis dissimilarities equal to 1, i.e., perfect dissimilarity, motivates a generative model that can accommodate point masses at the boundary of the unit interval.

\subsection{A brief review of GDMs}

Generalized dissimilarity modeling (GDM) treats pairwise ecological dissimilarities between sites as the response and relates them to differences in environmental and geographic covariates. Specifically, a GDM specifies a transformed \emph{ecological} distance as a sum of absolute differences in monotonically warped covariates, where each covariate is passed through a smooth, monotone function typically specified using I-spline basis expansions. This warping allows for flexible, nonlinear relationships between environmental gradients and compositional turnover. The transformed mean is then mapped to the unit interval via a link function (e.g., $\mu = 1 - e^{-\eta}$), and model fitting proceeds using iteratively reweighted least squares (IRLS) with weights proportional to $\mu(1-\mu)$, analogous to a binomial variance structure.

While GDM has become a widely used tool for analyzing beta diversity \citep[see][and references therein]{ferrier2007using,Mokany2022-dr}, its statistical foundation remains limited \citep[again, see][]{white2024generative}. Most notably, GDM does not arise from a formal likelihood, instead, as above, relying on an ad hoc combination of link and variance assumptions with subsequent fitting  via IRLS. This raises concerns about interpretability and validity, particularly given that pairwise dissimilarities exhibit clear dependence structures that are ignored in the fitting procedure. Additional issues include the use of a binomial-type variance function without a clear generative justification, the inability to properly quantify uncertainty due to violated independence assumptions in bootstrapping, and the lack of accommodation for the substantial incidence of ``1''s, again, complete dissimilarity, often observed in practice. More fundamentally, the GDM framework is not generative and cannot coherently produce the dissimilarity data it seeks to model, limiting its utility for inference, prediction, and model comparison.

To address these limitations, \cite{white2024generative} introduced the spatial generalized dissimilarity mixed model (spGDMM), within a fully generative, hierarchical Bayesian framework for modeling beta diversity. The spGDMM retains the appealing feature of monotone environmental warping while embedding it within a coherent probabilistic model that enables full likelihood-based inference. Key advances include the introduction of spatial random effects to capture dependence among dissimilarities, flexible variance specifications, and a principled treatment of one-inflation through a latent variable construction. \cite{white2024generative} demonstrate that this framework yields improved predictive performance and enables uncertainty quantification, spatial interpolation, and model comparison within a unified setting. Beyond these core contributions, the spGDMM framework can naturally be extended, and we use it as our modeling foundation in this paper. 

\subsection{Our Contributions}\label{sec:contributions}

Several important aspects of beta diversity remain unaddressed by the spGDMM formulation \citep{white2024generative}. In particular, that work focused on purely spatial dissimilarities and did not consider temporal evolution or joint modeling across time points. As a result, it cannot capture between-year beta diversity or the dynamic evolution of environmental warping functions under changing ecological conditions. Further, while the framework accommodates continuous covariates, it does not explicitly address the incorporation of categorical or ordinal predictors within the dissimilarity structure.

In this paper, we offer a joint dynamic spatiotemporal extension of the spGDMM to address these gaps, enabling joint modeling of beta diversity across space and time, accommodating more general covariate types, and allowing the effects of environmental drivers on dissimilarity to evolve dynamically. We call this approach the stGDMM (spatial and temporal generalized dissimilarity mixed modeling). The primary advantage of our approach is that by including between-year dissimilarities we more than double the amount of data relative to a year by year study of beta diversity. Specifically, the ${100 \choose  2}= 4950$ dissimilarities arise from $2450$ within-year beta dissimilarities (1225 for each year), $50$ within-site and between-year dissimilarities, and 2450 between-year and between-site dissimilarities.

To accommodate this richer dataset, we include at least four elements novel to dissimilarity modeling. First, we include a global between-year difference that accounts for the potential that between-year differences may be higher than within-year differences. Second, we develop evolving distance and covariate warping functions that allow environmental effects to evolve. Third, we incorporate categorical variables into the dissimilarity modeling framework. Lastly, we introduce novel space-time random effects to account for residual structure in space and time not captured by environmental variables.

In the context of \cite{heino2024navigating}'s conceptual framework of beta diversity, these modeling components can be viewed as targeting distinct spatial and temporal features of compositional turnover. Spatial beta diversity is represented by the within-year, between-site dissimilarities, and is modeled through geographic distance, environmental and categorical covariate differences, and the spatial random effects for each year. Temporal beta diversity is represented by same-site, between-year dissimilarities, and is captured by the global between-year effect, the time-varying covariates such as years since last fire, and differences between co-located space-time random effects.  Temporal variation in spatial beta diversity is accommodated by allowing the warped covariate and distance effects to change across years through. This allows the relationship between environmental gradients and spatial turnover to evolve through time. Finally, spatial variation in temporal beta diversity is modeled through the space-time random effect, which allows temporal change in community composition to vary across the landscape after accounting for measured environmental drivers. Thus, the proposed model links \cite{heino2024navigating}'s four conceptual aspects of beta diversity to explicit components within a single generative model for pairwise dissimilarities on $(t,\bs)\times(t',\bs')$.

In Section \ref{sec:data}, we describe the data used in this study and explore preliminary summaries that motivate our modeling decisions. In Section \ref{sec:methods}, we present the modeling framework and model comparison approaches. Section \ref{sec:results} shows results, including model comparison, estimated dynamic environmental effects, and spatiotemporal random effects. Lastly, in Section \ref{sec:summary} we summarize the paper and discuss potential avenues for future research.

\section{Dataset and Exploratory Analysis}\label{sec:data}

\subsection{Data Description}

\begin{figure}[h]
\begin{center}
    
        \includegraphics[width=0.54\textwidth]{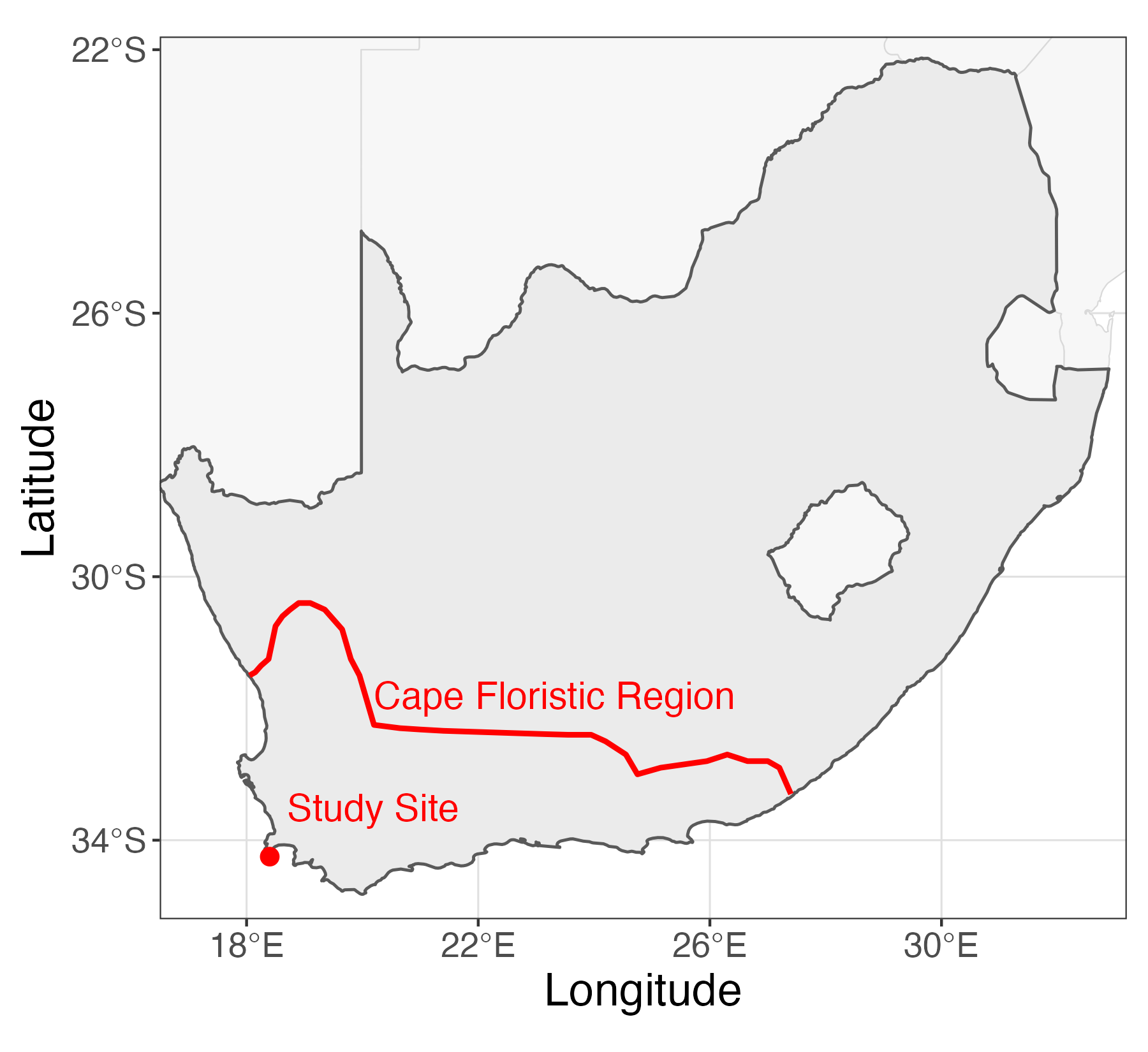}
    \includegraphics[width=0.45\textwidth]{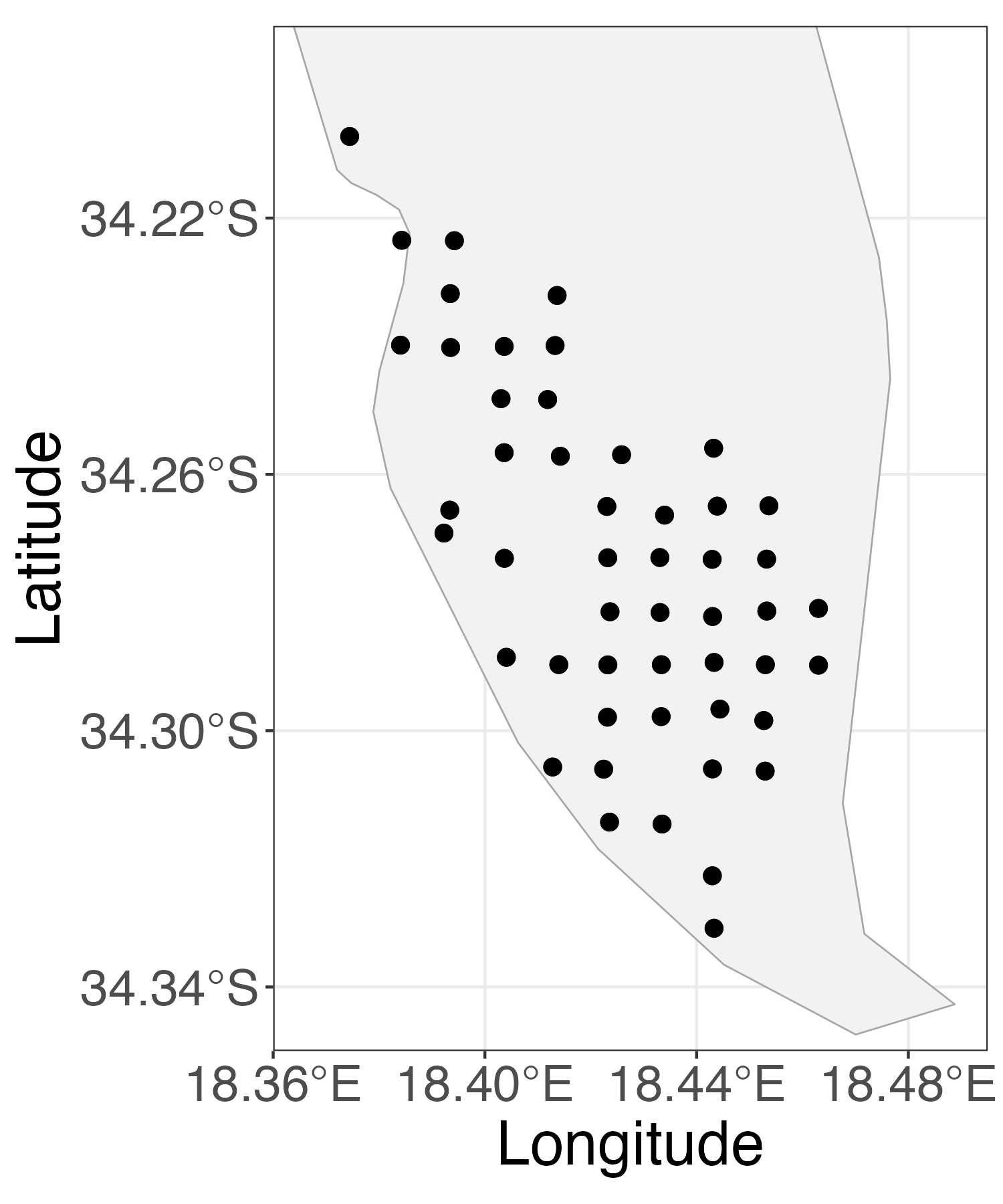}
    \end{center}
\caption{Study region and sampling locations. The left panel shows the location of the study area within South Africa, and the right panel shows the 50 sampling sites in the local study region within the Cape of Good Hope section of Table Mountain National Park. The study extent is geographically compact, with the maximum distance between sites equal to 15.1 km.}\label{fig:map}
\end{figure}

We consider fifty 5 by 10 m$^2$ permanent relevés (sites) surveyed in 1996 and 2021 (25 years apart) in the Cape Floristic Region of South Africa (Figure \ref{fig:map}). We consider each of the 100 unique site-year combinations as an observation. While the largest distance between sites is only 15.1 km, the study area contains over 1500 vascular plant species and the Cape Floristic Region is well known for its extremely high floristic turnover \citep{slingsby2017intensifying}. 

For simplicity, we denote the two times as $t = 1, 2$. The large temporal separation between the two surveys is both powerful and limiting. This gap of over 25 years enables us to address long-term questions about climate dynamics and their effects on beta diversity in this ecosystem. However, this large temporal separation limits finer scale ecological inference and statistical modeling.


In Section \ref{sec:summary}, we discuss more general temporal modeling, in particular both continuous and autoregressive discrete time specifications. While these specifications may apply to certain ecological datasets, regions are often only revisited sporadically and irregularly in time, so the need for such modeling will depend on the application.

In beta diversity modeling with temporal replication, we define dissimilarity between $\bY_t(\bs)$ and $\bY_{t'}(\bs')$, where each denotes a vector of species abundances or presence-absence indicators at time-location pairs $(t,\bs)$ and $(t',\bs')$, respectively. Across the 100 site-years in this study, we observed 360 unique plant species. Species richness differed between survey years, with 315 species observed in 1996 and 256 species observed in 2021. At the site-year level, richness ranged from 11 to 59 species, with a median of 35 species per site-year. Because the 50 sites were surveyed at two time points, the analysis includes 100 site-years and ${100 \choose 2} = 4950$ pairwise dissimilarities.

Here, we adopt the widely used Bray-Curtis (BC) dissimilarity,
\begin{equation}
Z_{t,t'}(\bs,\bs') = \frac{\sum_j | Y_{t}^{(j)}(\bs) - Y_{t'}^{(j)}(\bs')|  }{\sum_j | Y_{t}^{(j)}(\bs) + Y_{t'}^{(j)}(\bs')|},
\end{equation}
where $Y_{t}^{(j)}(\bs)$ is the abundance or presence of the $j$th species at site $\bs$ and time $t$. In this analysis, we treat the data as presence-absence data.

When $Z(\bs, \bs') = 1$, sites $\bs$ and $\bs'$ have no common species. If $Z_{t,t'}(\bs, \bs') = 0$, then this implies that $Y_{t}^{(j)}(\bs) = Y_{t'}^{(j)}(\bs')$ for all $j$.  Incidence of  $Z_{t,t'}(\bs, \bs') = 0$ will occur when $Y_{t}^{(j)}(\bs)$ records presence/absence of species $j$ at site $\bs$, perhaps when $\bs= \bs'$ with $t \neq t'$. There are no instances of 0 dissimilarity in the dataset; however, zeros can be quite common in some biomes \citep[see][for a discussion about Arctic biomes]{garcia2025plant}.  In our data, there are 151 dissimilarities (3.1\%) exactly equal to 1, i.e., complete dissimilarity\footnote{With a larger spatial range we see a much greater incidence of $1$'s.  For example \cite{white2024generative} found nearly $50\%$ $1$'s for the BC index in a different part of the Cape Floristic Region.}

As is customary in beta diversity modeling, environmental predictors were selected to explain compositional dissimilarity among sites across space and time in the study region. Most of the variables in this analysis are  continuous and purely spatial, i.e., static through time but vary spatially across sampling locations.  These include elevation, slope, long-term July minimum average temperature, and soil moisture class. 
These variables summarize topographic, long-term thermal, and hydrological heterogeneity that may influence species turnover by altering local habitat suitability, microclimate, and/or resource availability. 

However, we have one variable in our dataset, time since last fire, which varies both spatially and temporally, reflecting a dynamic disturbance history at each site. Fire history is particularly relevant in fire-prone systems, where post-fire succession and recovery can strongly influence community composition \citep{slingsby2017intensifying,verboom2024fire}. 

In addition to environmental differences, we use geographic distance between sites as an explanatory variable. This term allows the model to capture spatial structure in dissimilarity that may arise from dispersal limitation, unmeasured environmental gradients, or residual spatial autocorrelation. These variables are described in Table \ref{tab:covariates}.  

\begin{table}[h]
\centering
\scriptsize
\caption{Covariates used to model pairwise community dissimilarity. Environmental covariates were warped to allow nonlinear relationships with beta diversity.}
\label{tab:covariates}
\begin{threeparttable}
\begin{tabular}{llll}
\hline
\textbf{Covariate} & \textbf{Variable Type} & \textbf{Range} & \textbf{Ecological Meaning} \\
\hline
Elevation (m) & Spatial & 14.4--223.9 & Topographic position \\
Slope (grade) & Spatial & 0.006--0.296 & Terrain steepness \\
Average July minimum temperature ($^\circ$C) & Spatial & 9.10--10.33 & Winter thermal conditions \\
Moisture class & Spatial & 1--4\tnote{a} & Ordinal moisture gradient \\
Years since last fire & Spatial, temporal & 3--46 & Disturbance and succession \\
Distance between sites (km) & Pairwise spatial & Site-pair specific & Geographic separation \\
\hline
\end{tabular}

\begin{tablenotes}
\footnotesize
    \item[a] Moisture classes include: 1 = Seasonal wetland (shallow water table, occasionally inundated), 2 = Seasonally moist but never inundated, 3 = Damp landscape position, 4 = Well-drained.
\end{tablenotes}
\end{threeparttable}
\end{table}

Beta diversity modeling typically explains between-site dissimilarity by warping continuous environmental variables. These warping functions allow the effect of each covariate on beta diversity to be nonlinear, so that equivalent changes in a raw environmental variable need not correspond to equivalent changes in community dissimilarity across the full range of that variable. 

In addition to the continuous variables, our study includes an ordinal variable denoting soil moisture class. To include ordinal variables within the beta diversity modeling framework requires care, which we discuss in Section \ref{sec:methods}.

Again, the unique aspect of our data is the temporal component, and this allows us to explore whether beta diversity behaves differently over time. We evaluate whether values differ overall, in variability, or through the effects of environmental variables over time. Specifically, for all of the predictor variables described in Table \ref{tab:covariates}, we explore whether the effect of these variables is different for the two time periods (Section \ref{sec:methods}). Capturing these features is at the core of our methodological contribution.

\subsection{Exploratory Data Analysis}

\begin{figure}[h]
    \centering
    \includegraphics[width=\textwidth]{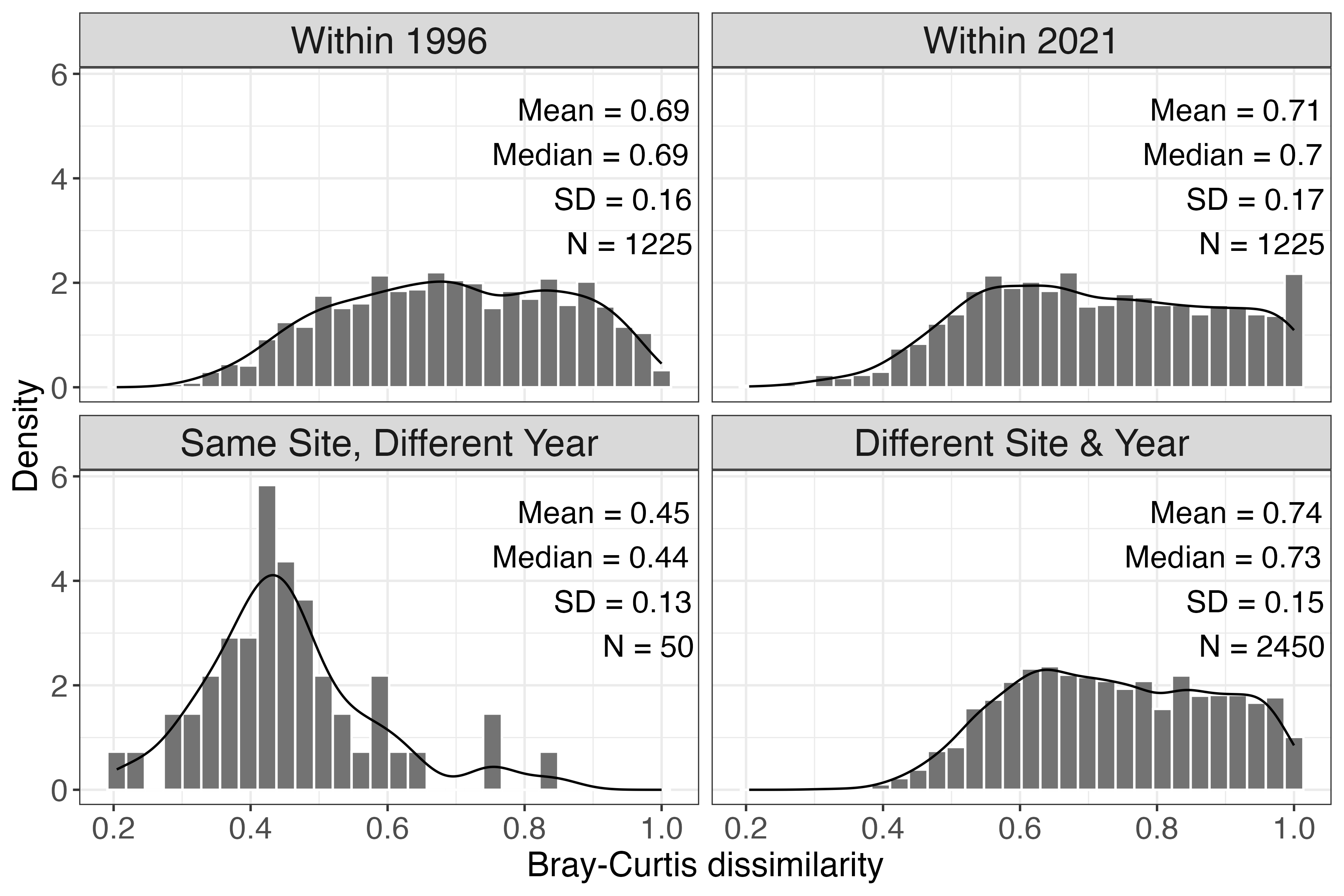}
\caption{Empirical distributions of Bray--Curtis dissimilarities by temporal comparison. Histograms show pairwise dissimilarities between sites surveyed in 1996, between sites surveyed in 2021, and between site-years from different survey years. Each panel reports the empirical mean, standard deviation, and sample size.}\label{fig:hist_BC}
\end{figure}

Here, we explore the dataset to highlight attributes that motivate future modeling choices. 
In Figure \ref{fig:hist_BC}, we plot histograms of BC dissimilarities, where the dissimilarities come from (i) spatial differences among plots in 1996, (ii) spatial differences among plots in 2021, (iii) the same site at different years, and (iv) spatial and temporal differences among plots between years. These histograms also include empirical summaries. Turnover values from the same site at different years have a mean of 0.45. Dissimilarities from spatial differences from among plots in 1996 are generally lower than the spatial differences from plots among plots in 2021 (0.69 vs. 0.71). The average dissimilarity is highest (0.74) when the plots come from different years and plots. 


\begin{figure}[h]
    \centering
    \includegraphics[width=0.5\textwidth]{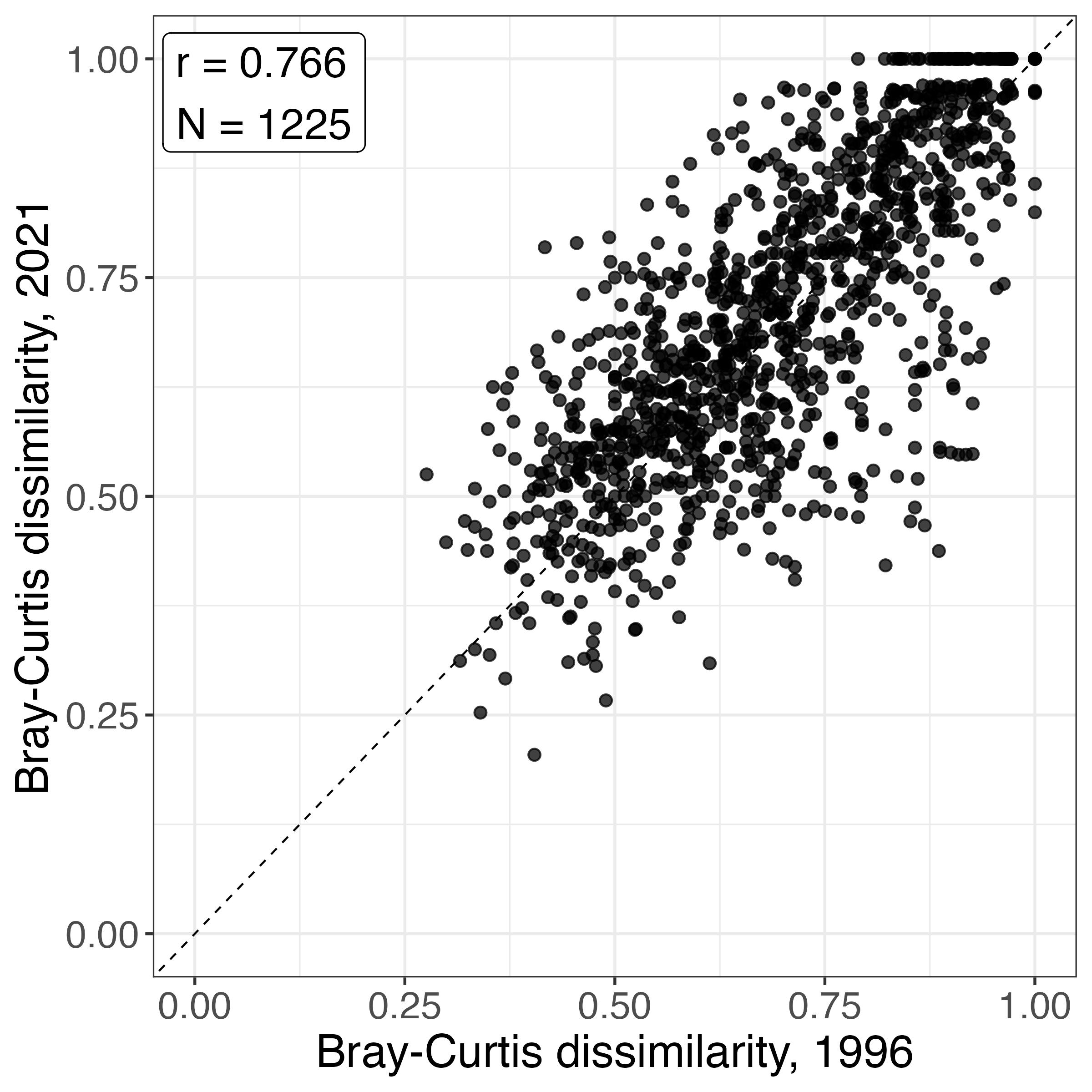}
\caption{Temporal persistence in pairwise Bray--Curtis dissimilarity. Each point represents one pair of sampling sites, with the x-axis showing the dissimilarity between the two sites in 1996 and the y-axis showing the dissimilarity for the same pair of sites in 2021. The dashed line is the one-to-one line.}\label{fig:scatter_line}
\end{figure}

To assess temporal persistence in community dissimilarity, we compared Bray-Curtis dissimilarities for the same pairs of sampling locations across the two survey years. Each point in Figure \ref{fig:scatter_line} represents a pair of sites, with the x-axis giving the pairwise dissimilarity in 1996 and the y-axis giving the dissimilarity for the same pair in 2021. The Pearson correlation between paired dissimilarities was $r = 0.766$ across N = 1225 site pairs (equivalently, $R^2 = 0.591$), suggesting substantial but imperfect temporal persistence in the spatial structure of beta diversity. A similar result was shown by \cite{thuiller2007stochastic} for the same plots for the years 1966 and 1996.

This pattern motivates the use of a modeling framework that can account for dependence in pairwise dissimilarities across both space and time, rather than treating the observed dissimilarities as independent or purely contemporaneous responses. However, these patterns also suggest the need for the model to evolve over time. These temporal differences could be explained by a variety of factors: overall shifts between the time periods, time-dependent environmental effects, or other more complex space time patterns not identified.

\section{Methods and Models}\label{sec:methods}

\subsection{Joint spatial and temporal generalized dissimilarity mixed model}\label{sec:stgdmm}


We extend the Bayesian spatial approach for generalized dissimilarity modeling in \cite{white2024generative} to account for dynamics. We refer to this model as a joint dynamic spatio-temporal generalized dissimilarity mixed model (stGDMM), specified as follows. Though we have only two time points in our application, for notational convenience, we continue to index time by $t$ and $t'$.

Because of the potential occurrence of $Z_{t,t'}(\bs, \bs') = 0$ and $Z_{t,t'}(\bs, \bs') = 1$, we account for potential observed point masses at $0$ and $1$  through a latent variable $V_{t,t'}(\bs, \bs')$, defined as
\begin{equation}
    Z_{t,t'}(\bs, \bs') = \begin{cases}
     0 &\text{if }  V_{t,t'}(\bs, \bs') \leq 0, \\
       V_{t,t'}(\bs, \bs')  & \text{if } V_{t,t'}(\bs, \bs') \in (0,1),\\
         1 & \text{if }  V_{t,t'}(\bs, \bs') \geq 1.
    \end{cases}
\end{equation}


Next, we specify the model for $V_{t,t'}(\bs, \bs')$ in what is a customary geostatistical form.  That is,
\begin{equation}\label{eq:v_mod}
V_{t,t'}(\bs,\bs')
=
\underbrace{\mu_{t,t'}(\bs,\bs')}_{\text{Environmental Mean}}
+
\underbrace{\eta_{t,t'}(\bs,\bs')}_{\text{Space-Time Random Effect}}
+
\underbrace{\epsilon_{t,t'}(\bs,\bs')}_{\text{Pure Error}},
\end{equation}
where $\epsilon_{t,t'}(\bs, \bs') \overset{ind}{\sim} \mathcal{N}(0,\sigma_{t,t'}^2).$ Thus, we assume that dissimilarities across different times and between times are heteroscedastic. 

The key feature of GDM modeling is to introduce \emph{warping} into the discrepancy between the effects of environmental features at $(\bs,t)$ and $(\bs',t')$.  That is, the transformed environmental mean is not specified linearly in these differences. A general form for the environmental mean is
\begin{equation}\label{eq:mu_mod}
    \begin{aligned}
        \mu_{t,t'}(\bs, \bs') &= \beta_0  &\text{Intercept} \\ 
        &+ \delta \bone_{t \neq t'}(t,t')  &\text{Global Time Effect} \\
        &+ h_{t,t'}(\|\bs - \bs' \|) &\text{Warped Distance} \\
        &+ \sum_{k = 1}^K | g_{k,t}(X_{k,t}(\bs)) - g_{k,t'}(X_{k,t'}(\bs')) |  &\text{Time-Specific Continuous Warpings} \\
        &+ | D_{t}(\bs)^\top\bgamma_t - D_{t'}(\bs')^\top\bgamma_{t'}|  &\text{Time-Specific Ordinal Effects}. \\
    \end{aligned}
\end{equation}


While the model form in expression \eqref{eq:v_mod} is general over time, the expression in \eqref{eq:mu_mod} reminds us that, in our data application, we are working with only two time points. 
We include a global intercept, and we include a single $\delta$ to capture the effect of the change in time. We discuss more general specifications in Section \ref{sec:summary}. Further, again with two time points, we use the general decomposition of $h_{t,t'}(\cdot)$ and $g_{k,t}(\cdot) = \alpha_{k,t} F_{k,t}(\cdot)$, $t=1$ or $t=2$,  as in \cite{white2024generative}, where $F_{k,t}(\cdot)$ is a random CDF and $\alpha_{k,t}$ is a variable importance parameter.  This enables convenient assessment of the relative importance of the regressors under the warping.  

Following \cite{ferrier2007using,white2021hierarchical,white2024good}, we use cubic I-spline basis functions to model $F(\cdot)$ \citep{ramsay1988monotone}, each with two interior knots at the 33rd and 67th percentiles of the predictor. Thus, $$F_{k,t}(x) = \sum_{j = 1}^{5} I_{k,t,j}(x) \beta_{k,t,j} ,$$
where $I_{k,t,j}$ are I-spline basis functions and $\sum_{j = 1}^{5} \beta_{k,t,j} = 1$ to ensure that $F_{k,t}(x)$ is a CDF. 


To ensure monotonicity with ordinal categorical variables, soil moisture class in our case, we constrain $\bgamma_t \geq 0$ and specify dummy variables to enable cumulative effects. For example, we show the specification of $D_{t'}(\bs')$ for all observed values of soil moisture class (1-4) below:
\begin{equation*}
    \begin{aligned}
        &\text{If Soil Moisture Class} = 1,\text{ then }  D_{t}(\bs) = (0,0,0)^\top; \\
                &\text{If Soil Moisture Class} = 2,\text{ then } D_{t}(\bs) = (1,0,0)^\top;\\
        &\text{If Soil Moisture Class} = 3,\text{ then } D_{t}(\bs) = (1,1,0)^\top;\\
        &\text{If Soil Moisture Class} = 4,\text{ then } D_{t}(\bs) = (1,1,1)^\top.\\
    \end{aligned}
\end{equation*}
This cumulative specification of the ordinal dummy variables, as well as the non-negative constraint on $\bgamma_t$, ensures that the soil moisture class is appropriately ordered within the model. Nominal categorical variables could be handled similarly but without the cumulative specification of the dummy variable. 


Again, with two time points, the spatiotemporal random effects for stGDMM are specified through a bivariate Gaussian Process (GP) using a separable cross-covariance structure 
\begin{equation}
    \begin{aligned}
        \eta_{t,t'}(\bs, \bs') &= \left( \psi_t(\bs) - \psi_{t'}(\bs') \right)^2, \\
  \bpsi(\bs) = \begin{pmatrix}
       \psi_{1}(\bs) \\ \psi_2(\bs)
   \end{pmatrix} &\sim \mathcal{GP}\left( 
   \begin{pmatrix} 0 \\ 0\end{pmatrix}, \bV e^{-\frac{\|\bs - \bs' \|}{\rho}} \right), \\
    \end{aligned}
\end{equation}
where $\bV$ is an unstructured $2\times2$ covariance matrix modeling between-year process correlation for $\psi$. The random effects here are novel in that they operate on $(t, \bs) \times (t', \bs')$ space, making them difficult to display and interpret.  Because of the lack of identifiability of scale and range parameters for Mat\'ern Gaussian process models \citep{zhang2004inconsistent}, we fix $\rho = 1.51$ km (one-tenth of the maximum distance between sites) so that $\bV$ is identified and can be well-estimated.


Lastly, with regard to the error term $\epsilon_{t,t'}(\bs,\bs')$, we use a heteroscedastic error term motivated in part by our exploratory analysis in Section \ref{sec:data}. When $t = t' = 1$, $\text{Var}(\epsilon_{t,t'}(\bs,\bs')=\sigma^{2}_{11}$, while $\text{Var}(\epsilon_{t,t'}(\bs,\bs')=\sigma^{2}_{22}$ when $t = t' = 2$. Lastly, when $t \neq t'$, $\text{Var}(\epsilon_{t,t'}(\bs,\bs')=\sigma^{2}_{12}$. This allows the model to capture evolving variability, as well as a different variance for between-year dissimilarities.

A key modeling question is whether the inclusion of the space by time random effects is justified. Including a GP offers the possibility of explaining latent dependence in the data, but it also introduces substantial model complexity. Thus, we use out-of-sample prediction (described in Section \ref{sec:mod_comp}) to determine whether the added complexity is justified by improved model performance. 

Another modeling question is how the effects associated with the environmental variables evolve? For us, does the overall strength of environmental effects change between the two time points considered? Does the shape of the environmental variable effect (warping) change over time? 

To answer these questions regarding environmental dynamics, we consider comparison of four specifications of $g_{k,t}(\cdot)$:
\begin{enumerate}
    \item $g_{k,t}(\cdot) = \alpha_{k} F_{k}(\cdot)$ - shared warping functions across time.
    \item $g_{k,t}(\cdot) = \alpha_{k,t} F_{k}(\cdot)$ - shared shape across time, but different variable importance by time.
    \item $g_{k,t}(\cdot) = \alpha_{k} F_{k,t}(\cdot)$ - shared variable importance, but unique shape each year.
    \item $g_{k,t}(\cdot) = \alpha_{k,t} F_{k,t}(\cdot)$ - unique warping functions across year.
\end{enumerate}
As will be done with the space-time random effects, these modeling questions will be answered in terms of out-of-sample predictive performance.



\subsection{Priors, Model Fitting, and Prediction}

We assign weakly informative prior distributions to model parameters, using the scale and constraints of the data to guide their specification. In particular, because Bray--Curtis dissimilarities are bounded on $[0,1]$, prior distributions for parameters governing the latent mean and variance need not place substantial mass on values that imply extreme or implausible dissimilarities. Thus, our priors are weakly informative in the sense that they allow us to estimate a broad range of relationships between environmental covariates and dissimilarity. 

In our model, we constrain the I-spline basis coefficients for each covariate and year to sum to one ($\sum_{j = 1}^{5} \beta_{k,t,j} = 1$). To impose this constraint, we define $$\beta_{k,t,j} = \frac{\beta^*_{k,t,j}}{\sum_{j = 1}^{5} \beta^*_{k,t,j} }.$$
To ensure parameter identifiability under this constraint, we fix $\log(\beta_{k,t,1}^*) = 0$ for all $k$ and $t$.

We assume the following prior distributions:
$\beta_0 \sim \text{Normal}\left(0.5,1\right)$, 
$\log(\delta) \sim  \text{Normal}\left( -1.5,1 \right)$, 
$\log(\alpha_{k,t}) \sim \text{Normal}\left( -2, 1 \right),$ 
$\log( \beta^*_{k,t,j}) \sim  \text{Normal}\left( 0,3 \right),$ 
$\log(\gamma_{k,t}) \sim \text{Normal}\left( -1.5,1 \right)$, 
$\bV \sim \text{Inverse-Wishart}\left(3,0.1\mathbb{I}\right)$, and 
$\sigma^2_{t,t'} \sim \text{Inverse-Gamma}\left(3,0.03\right)$.

We fit the model using Markov chain Monte Carlo (MCMC). When full conditional distributions were available in closed form, we used Gibbs sampler updates. Otherwise, parameters were updated using Metropolis--Hastings steps within the Gibbs sampler. Because many parameters enter the model through nonlinear warping functions, latent truncation, or covariance structures, most updates were implemented using Metropolis--Hastings proposals. The model was not computationally expensive for this dataset. Thus, for each model fit, we ran the sampler for 300,000 iterations, discarded the first 100,000 iterations as burn-in, and retained every 20th draw thereafter, yielding 10,000 posterior samples for inference and prediction.

We use each of the retained posterior samples for all model parameters, $\btheta^{(1)},\ldots,\btheta^{(M)}$, to obtain posterior predictions for held-out or unobserved Bray--Curtis dissimilarities. Specifically, for a pair of site-years $(t,\bs)$ and $(t',\bs')$, the posterior predictive distribution is
\begin{equation}\label{eq:post_pred}
\left[ Z_{t,t'}(\bs,\bs') \mid \bZ\right]
=
\int
\left[Z_{t,t'}(\bs,\bs' )  \mid V_{t,t'}(\bs,\bs' \right]  \left[V_{t,t'}(\bs,\bs' )  \mid \btheta\right]
\left[\btheta \right]
\, d\bV d\btheta,
\end{equation}
which we approximate using composition sampling \citep{tanner1996}. That is, for each posterior draw $\btheta^{(m)}$, we compute the corresponding environmental mean and space-time random effect, simulate a latent value $V^{(m)}_{t,t'}(\bs,\bs')$ from the data model in \eqref{eq:v_mod}, and then transform this value to the observed dissimilarity scale using the censoring rule defining $Z_{t,t'}(\bs,\bs')$. This yields posterior predictive samples $Z^{(1)}_{t,t'}(\bs,\bs'),\ldots,Z^{(M)}_{t,t'}(\bs,\bs')$, which summarize uncertainty in the predicted dissimilarity for the site-year pair. These posterior predictive samples are used for out-of-sample prediction, uncertainty quantification, and the calculation of predictive scoring rules in the cross-validation analysis discussed in Section \ref{sec:mod_comp}.



\subsection{Model Comparison}\label{sec:mod_comp}

We assess out-of-sample predictive performance using 10-fold cross-validation, where folds were constructed at the site-by-time level rather than at the level of individual dissimilarities. In each fold, ten of the 100 unique site-year combinations are held out. All Bray--Curtis dissimilarities involving at least one held-out site-year combination are excluded from model fitting. In total, this means that ${100 \choose 2} - {90 \choose 2} = 945$ dissimilarities are held out in each fold. The model is then fit to the remaining ${90 \choose 2}$ dissimilarities, yielding posterior samples $\theta^{(1)},\ldots,\theta^{(M)}$. These posterior samples are used to generate posterior predictive samples $Z_{t,t'}^{(1)}(\bs,\bs'),\ldots,Z_{t,t'}^{(M)}(\bs,\bs')$ for the held-out dissimilarities $Z_{t,t'}(\bs,\bs')$. This design targets prediction for new or partially unobserved site-years, which we view as more relevant for the intended application than randomly withholding individual pairwise dissimilarities.

We summarize predictive performance using several complementary scoring rules. We use root mean square and mean absolute errors (RMSE and MAE) to evaluate accuracy of the posterior predictive mean. In addition, we use the continuous rank probability score (CRPS), to assess the full posterior predictive distribution \citep{brown1974,matheson1976,gneiting2007}. We estimate CRPS using the empirical CDF of posterior predictions \citep{kruger2021predictive},
\begin{equation}
\frac{1}{M} \sum_{j=1}^M \mid Z_{t,t'}^{(j)}(\bs, \bs') - Z_{t,t'}(\bs, \bs') \mid  - \frac{1}{2M^2} \sum_{j=1}^M \sum_{k=1}^M \mid  Z_{t,t'}^{(j)}(\bs, \bs') - Z_{t,t'}^{(k)}(\bs, \bs') \mid.
\end{equation}


We also use logarithmic scores to measure the predictive density assigned to the held-out observations \citep{good1952rational,gneiting2007}. Because $Z_{t,t'}(\bs,\bs')$ has a mixed distribution, with point masses at 0 and 1 and a continuous density on $(0,1)$, the logarithmic score must account for both the discrete and continuous components of the posterior predictive distribution. For each held-out value, we compute the negative logarithmic score (LogS) as
\begin{equation}\label{eq:logs}
\begin{cases}
-\log\left[\frac{1}{M} \sum_{m=1}^M\Phi\left\{\frac{0 - \mu^{(m)}_{t,t'}(\bs,\bs') - \eta^{(m)}_{t,t'}(\bs,\bs')}{\sigma^{(m)}_{t,t'}}\right\}\right],& \text{if } Z_{t,t'}(\bs,\bs') = 0, \\
-\log\left[\frac{1}{M} \sum_{m=1}^M\frac{1}{\sigma^{(m)}_{t,t'}}\phi\left\{\frac{Z_{t,t'}(\bs,\bs') - \mu^{(m)}_{t,t'}(\bs,\bs') - \eta^{(m)}_{t,t'}(\bs,\bs')}{\sigma^{(m)}_{t,t'}}\right\}\right],& \text{if } Z_{t,t'}(\bs,\bs') \in (0,1), \\
-\log\left[\frac{1}{M} \sum_{m=1}^M\left\{1 -\Phi\left(\frac{1 - \mu^{(m)}_{t,t'}(\bs,\bs') - \eta^{(m)}_{t,t'}(\bs,\bs')}{\sigma^{(m)}_{t,t'}}\right)\right\}\right],& \text{if } Z_{t,t'}(\bs,\bs') = 1.
\end{cases}
\end{equation}
Here, $\phi(\cdot)$ and $\Phi(\cdot)$ denote the probability density and distribution functions of a standard normal random variable. The Monte Carlo average is taken over the posterior predictive density or probability before applying the logarithm, so that \eqref{eq:logs} scores the full posterior predictive distribution rather than the conditional predictive distribution at a single posterior draw \citep{vehtari2017practical}. We compute \eqref{eq:logs} directly from the model-implied posterior predictive distribution rather than from an empirical approximation to the posterior predictive samples because (i) the distribution of $Z_{t,t'}(\bs,\bs')$ is mixed discrete-continuous and (ii) sample-based estimates of the logarithmic score can be unstable except under stringent conditions \citep{kruger2021predictive}.


\section{Results}\label{sec:results}

\subsection{Model Comparison Results}

Our primary model comparison is focused on models that differ in two ways: (i) inclusion or exclusion of space-time random effects and (ii) warping function specification. These model specifications are described in Section \ref{sec:stgdmm}. 

Warping Model 1 assumes that both covariate importance and warping shape are shared across years, whereas Warping Model 2 allows covariate importance to vary by year but keeps the warping shape fixed. Warping Model 3 instead keeps covariate importance fixed but allows the warping shape to vary, and Warping Model 4 allows both components to vary by year. We present the results of the model comparison in Table \ref{tab:model_comp}. 

Predictive performance is dramatically improved by including the proposed random effects in the model, suggesting a clear benefit of space-time modeling. On the other hand, predictive performance is broadly similar across the four warping specifications, indicating that, with only two time points, the main conclusions are not highly sensitive to the precise form of temporal sharing in the warping functions.

Nevertheless, Warping Model 3 achieves the best performance across RMSE, MAE, CRPS, and LogS. This model assumes $g_{k,t}(\cdot) = \alpha_k F_{k,t}(\cdot),$ so that the overall importance of covariate $k$ is shared across years, while the shape of the warping function is allowed to vary by year. In different words, the cross-validation results suggest that allowing temporal changes in the shape of the covariate effects improves prediction, while retaining a common variable-importance parameter provides useful regularization. Continuing, the fact that Warping Model 3 outperforms Warping Model 4 suggests that the fully year-specific specification may introduce unnecessary flexibility, without improving held-out predictive performance. Overall, these results support a model in which the environmental response surfaces may change over time, but the relative importance of the environmental covariates is stable across years. Because the model including space-time random effects and Warping, Model 3, is best in predictive performance, we present results for this model in what follows.

\begin{table}[h]
\centering
\caption{Model comparison from 10-fold cross-validation results for models with and without space-time random effects and that vary by four warping model specifications. Because we use the negative logarithm score, smaller values for all measures indicate better predictive performance. Best performances are printed in boldface.}\label{tab:model_comp}
\begin{tabular}{rr|rrrr}
  \hline
$\eta_{t,t'}(\bs, \bs')$ Included & Warping Model & RMSE & MAE & CRPS & LogS \\ 
  \hline
No &  1 & 0.1491 & 0.1241 & 0.0857 & -0.3998 \\ 
No &    2 & 0.1492 & 0.1242 & 0.0858 & -0.3984 \\ 
No &    3 & 0.1485 & 0.1238 & 0.0853 & -0.4051 \\ 
No &    4 & 0.1509 & 0.1258 & 0.0869 & -0.3865 \\ \hline
Yes &  1 & 0.1063 & 0.0831 & 0.0595 & -0.6933 \\
Yes &    2 & 0.1090 & 0.0856 & 0.0611 & -0.6694 \\ 
Yes &   3 & \textbf{0.1062} & \textbf{0.0823} & \textbf{0.0591} & \textbf{-0.7036} \\  
Yes &    4 & 0.1110 & 0.0870 & 0.0624 & -0.6365 \\ 
   \hline
\end{tabular}
\end{table}

\subsection{Environmental Predictor Effects}

\begin{figure}[h]
    \centering
        \includegraphics[width=\linewidth]{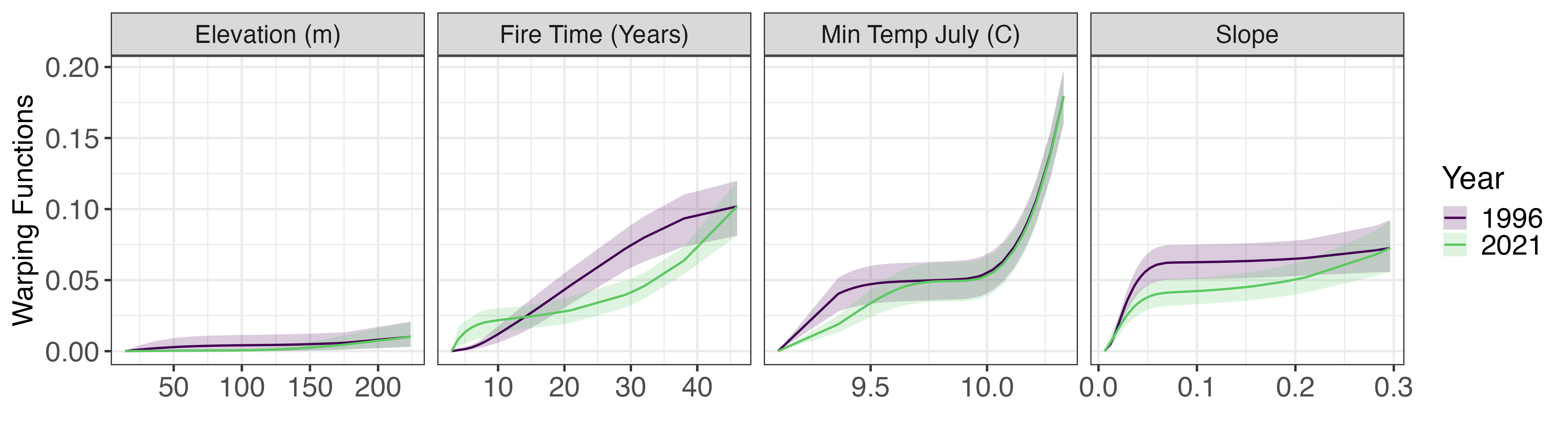}


\caption{Estimated warping functions under Warping Model 3. Curves show posterior mean warping functions for 1996 and 2021, and shaded bands show 90\% credible intervals. All panels use a common vertical scale, allowing comparison of the relative magnitudes of the warped covariate effects.}\label{fig:warping}
\end{figure}

\begin{figure}[h]
    \centering
 
        \includegraphics[width=0.67\linewidth]{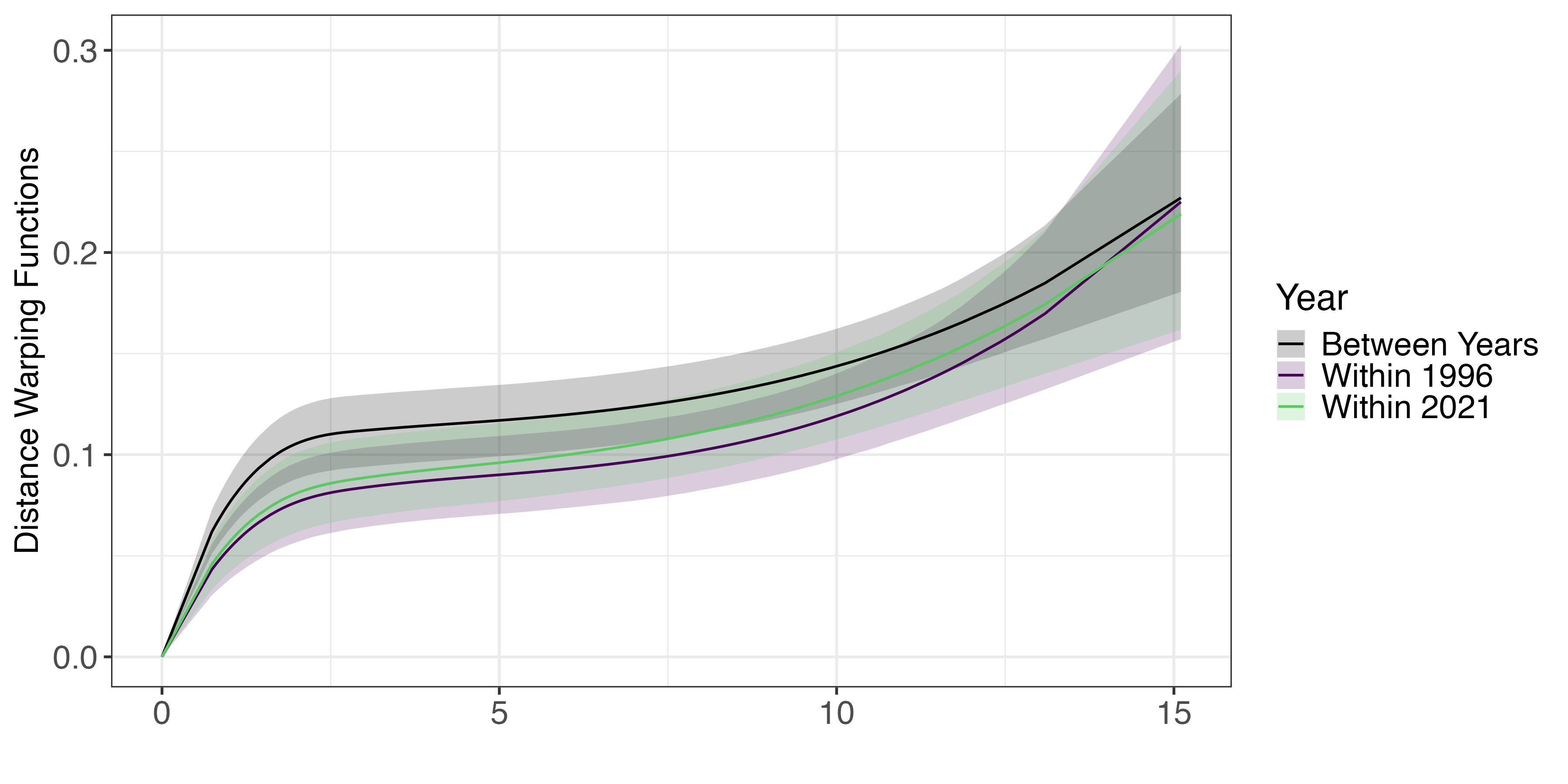}


\caption{Estimated warping functions for distance functions between years, within 1996, and within 2021. Curves show posterior mean warping functions for 1996 and 2021, and shaded bands show 90\% credible intervals.}\label{fig:dist_warping}
\end{figure}

We plot the posterior mean and 90\% credible interval for the warping functions of environmental variables under Warping Model 3, $g_{k,t}(\cdot) = \alpha_k F_{k,t}(\cdot),$ in Figure \ref{fig:warping}. Under the framework presented by \cite{heino2024navigating}, the differences between these curves represent how abiotic variables contribute to temporal variation in spatial beta diversity. Because all panels are shown on a common vertical scale, the relative magnitudes of the warped covariate effects can be compared across variables. The warping functions for distance, which are not constrained as the environmental variables, are plotted in Figure \ref{fig:dist_warping}.

Geographic distance exhibits the largest warped effect, increasing sharply at short distances and then continuing to increase more gradually over the observed range, suggesting that spatial separation is an important driver of compositional dissimilarity. In general, the distance warping functions between years is higher than those for plots within 1996 and 2021. This suggests that distance has greater effects on beta diversity for plots collected over different years than when plots were collected at the same time. The warping functions for July minimum temperature and fire time also show appreciable nonlinear effects, with relatively modest changes through much of the covariate range followed by steeper increases at larger values. In contrast, slope and elevation have smaller overall warped effects, although both show evidence of nonlinearity. Differences between years are present but generally modest. 

For all variables the 1996 and 2021 warping functions have broadly similar shapes, with some year-specific deviations in magnitude across portions of the covariate range. The 1996 functions tend to be slightly larger for slope, elevation, and intermediate-to-large fire times, whereas the 2021 functions are comparable or slightly larger at the upper end of the fire-time range and show a somewhat lower temperature effect over much of the observed range. Overall, the main covariate effects are qualitatively stable across years.  Allowing year-specific warping shapes allows us to capture moderate temporal differences in how environmental gradients are associated with beta diversity.

\begin{figure}[h]
    \centering
    \includegraphics[width=0.52\linewidth]{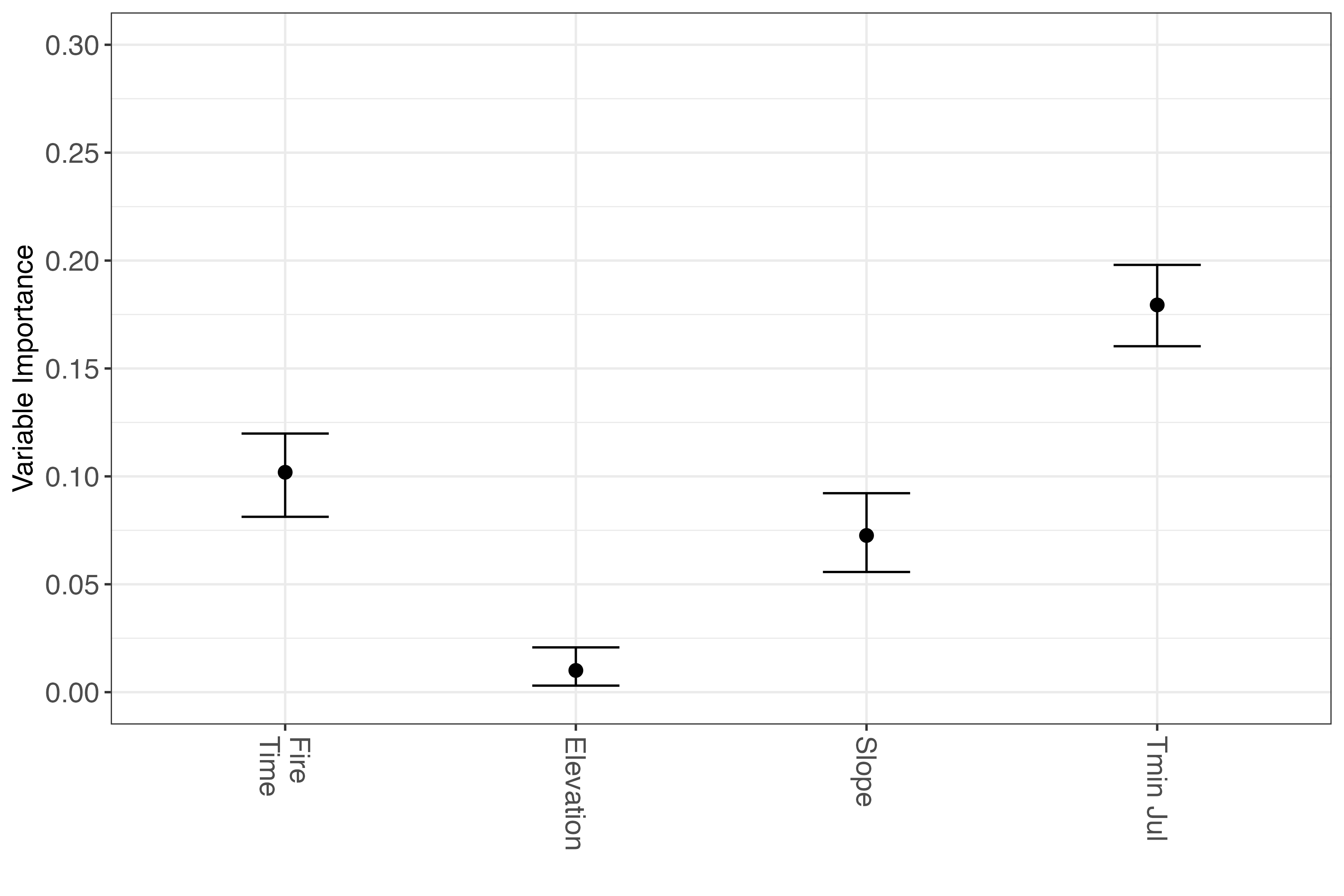}
        \includegraphics[width=0.47\linewidth]{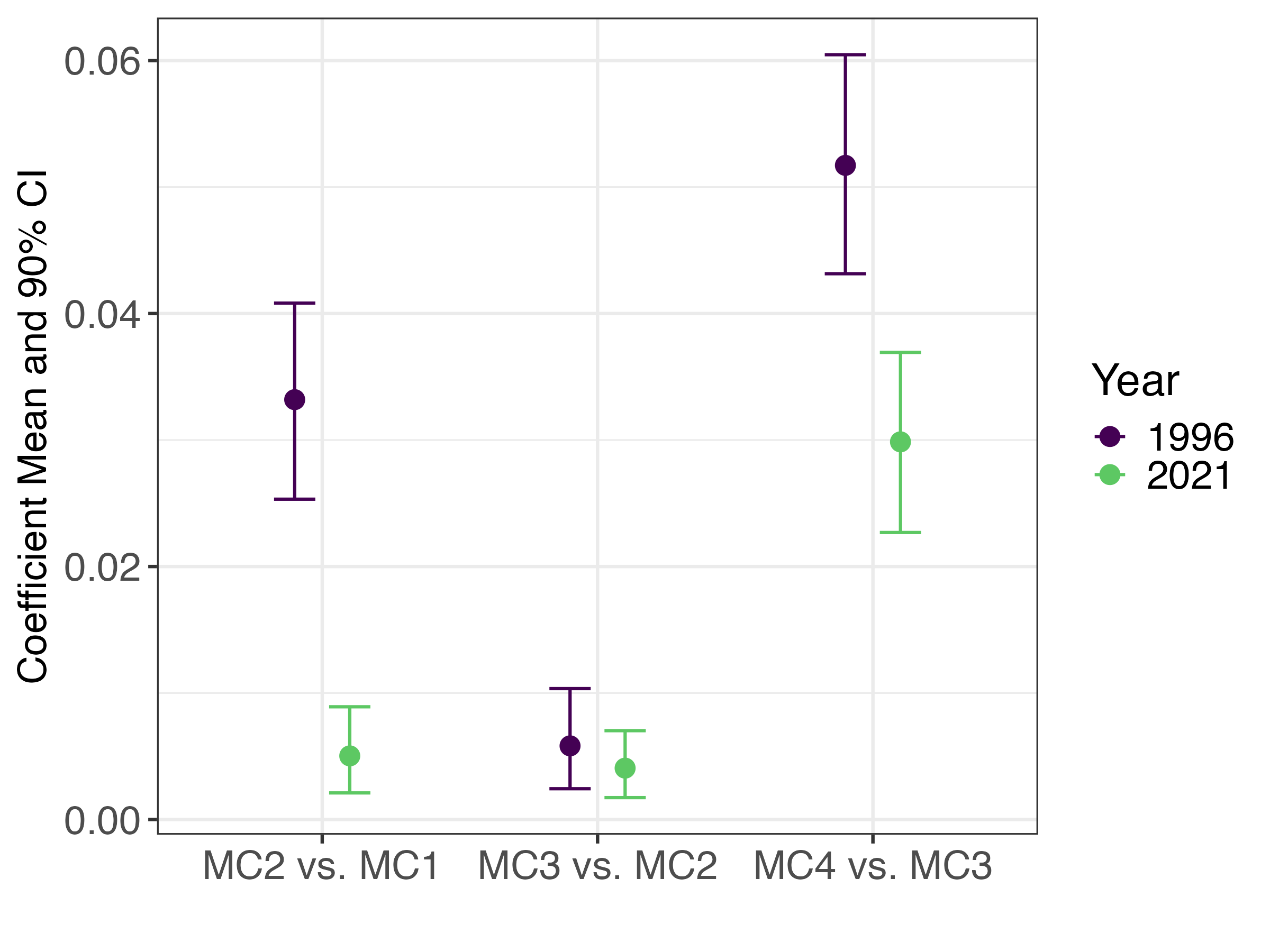}
\caption{Posterior summaries of covariate effects under Warping Model 3. (Left) Posterior means and 90\% credible intervals for the shared variable-importance parameters, $\alpha_k$, for continuous covariates and geographic distance. (Right) Posterior means and 90\% credible intervals for year-specific ordinal moisture class contrasts. The distance warping function is shared across years.} \label{fig:alpha}
\end{figure}

To separate the magnitude of each covariate contribution from the shape of its warped effect, Figure \ref{fig:alpha} summarizes the posterior distributions of the shared variable-importance parameters, $\alpha_k$, under Warping Model 3. Although the credible intervals indicate some posterior uncertainty, the ordering of the covariates is fairly clear: distance and minimum July temperature have the largest shared importance parameters, elevation has the smallest effect, while fire time and slope fall in between.

The ordinal moisture class effects suggest a stronger association between soil moisture conditions and community dissimilarity in 1996 than in 2021 (See Figure \ref{fig:alpha}). In particular, the estimated contrasts between adjacent moisture classes are consistently larger in 1996, especially for the transition from MC4 to MC3 and from MC2 to MC1. By contrast, the corresponding 2021 effects are smaller and have credible intervals closer to zero, indicating that differences among moisture classes contributed less to beta diversity in the more recent survey. This pattern suggests that soil moisture class was a more important axis of compositional differentiation in 1996, whereas contemporary community dissimilarity appears to be less strongly structured by this ordinal moisture gradient. However, these patterns are also influenced by changes in alpha diversity, i.e., local richness at each plot, and we observe lower species richness in 2021 compared to 1996. Losing species from plots can have a homogenizing effect, i.e., lower beta diversity, and thus plays an interactive role in making inferences about beta diversity patterns.


Table \ref{tab:param_post_summary} summarizes posterior inference for the intercept, global between-year effect, and the three error variance parameters. The posterior mean of $\delta$ is small and positive, suggesting that, after accounting for environmental covariates and random effects, between-year dissimilarities, i.e., temporal beta diversity,  are slightly larger than within-year dissimilarities on average, i.e., spatial beta diversity. We suggest that the $\delta$ term can be thought of as an indicator of temporal baseline shifts in turnover. Decreasing temporal beta diversity is caused by homogenization of communities while increasing beta diversity caused differentiation of communities \citep{rolls_biotic_2023}. In our modeling framework, if the $\delta$ term is large and negative, then the set of samples have become homogeneous from one time point to the next suggesting that some form of homogenization process is occurring. The causes of homogenization are varied but can include reductions in fire regime or other disturbances, prolonged species invasions, or altered land use (see \cite{rolls_biotic_2023} and references therein). Small values of $\delta$ could indicate the start of a trend or simply be within stochastic variation and the landscape is relatively stable. More time points should be analyzed to distinguish between growing trends and expected stochasticity. A large positive $\delta$ term would suggest a process that is greater differentiation of communities. Assuming measurements are made on time scale where speciation is not relevant, processes that could lead to strong positive $\delta$ include greater levels of localized disturbance or more fragmented cases of land use change. We also note that inference made with $\delta$ should also consider residual temporal parameter,  $\sigma_{12}^2$. A high value may indicate that certain unmeasured abiotic variables or biotic processes have not been included in the analysis and that inferences with $\delta$ should be made with caution. Some form of sensitivity analysis that explores different sets of model covariates is recommended in applied use cases. 

The posterior summaries for $\sigma_{11}^2$, $\sigma_{12}^2$, and $\sigma_{22}^2$ indicate modest differences in residual variability across the two within-year comparisons and the between-year comparison. In particular, the between-year residual variance, $\sigma_{12}^2$, is slightly smaller than the residual variances for the two within-year comparisons. Assuming that the abiotic variables included in the model adequately capture all the relevant factors related to dissimilarity, ecologists may consider the residual posteriors as indicating possible unmeasured biotic processes, e.g., invasions, disease, competition, and shifts in demography. While the residuals in the current example are rather small, one may interpret large differences in the estimates of  $\sigma_{11}^2$ and  $\sigma_{22}^2$ as an indicator that certain functional aspects of the community have shifted between the two time points because of biotic processes, e.g., perhaps the population of plants with longer distance dispersal mechanisms decreased between the two time points as the results of the introduction of an invasive specie.

\begin{table}[h]
\centering
\caption{Posterior summaries for selected model parameters. The 90\% credible interval is given by the 5th and 95th posterior quantiles.}
\label{tab:param_post_summary}
\begin{tabular}{lrrrr}
\hline
\textbf{Parameter} & \textbf{Mean} & \textbf{Median} & \textbf{SD} & \textbf{90\% Credible Interval} \\
\hline
$\beta_0$       & 0.4306 & 0.4306 & 0.0110 & $(0.4123,\ 0.4488)$ \\
$\delta$        & 0.0213 & 0.0212 & 0.0055 & $(0.0124,\ 0.0306)$ \\
$\sigma_{11}^2$ & 0.0104 & 0.0104 & 0.0004 & $(0.0097,\ 0.0112)$ \\
$\sigma_{12}^2$ & 0.0075 & 0.0075 & 0.0002 & $(0.0072,\ 0.0079)$ \\
$\sigma_{22}^2$ & 0.0098 & 0.0098 & 0.0004 & $(0.0091,\ 0.0105)$ \\
\hline
\end{tabular}
\end{table}


\subsection{Dynamic Spatial Random Effects}

\begin{figure}[h]
    \centering
    \includegraphics[width=0.32\linewidth]{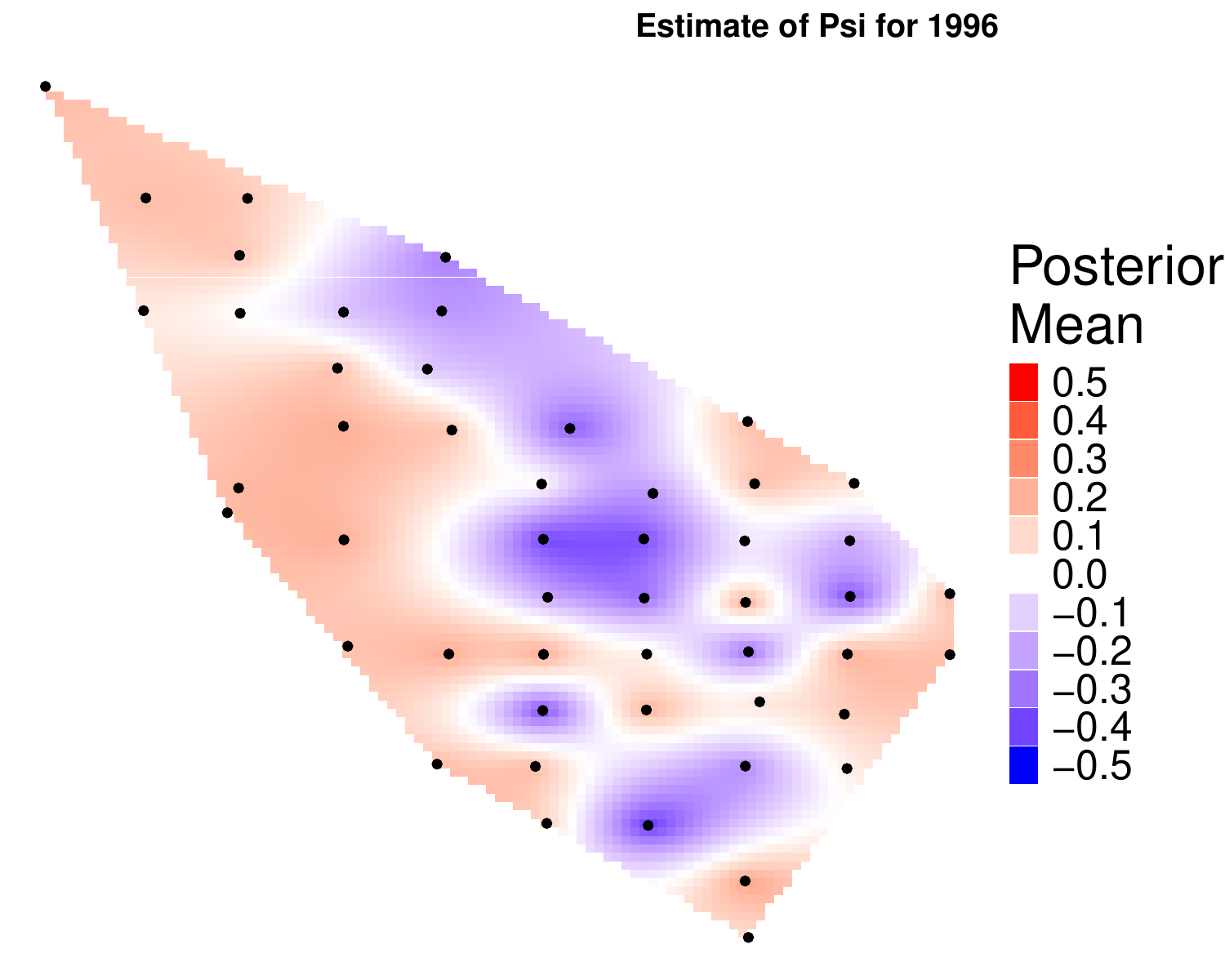}
    \includegraphics[width=0.32\linewidth]{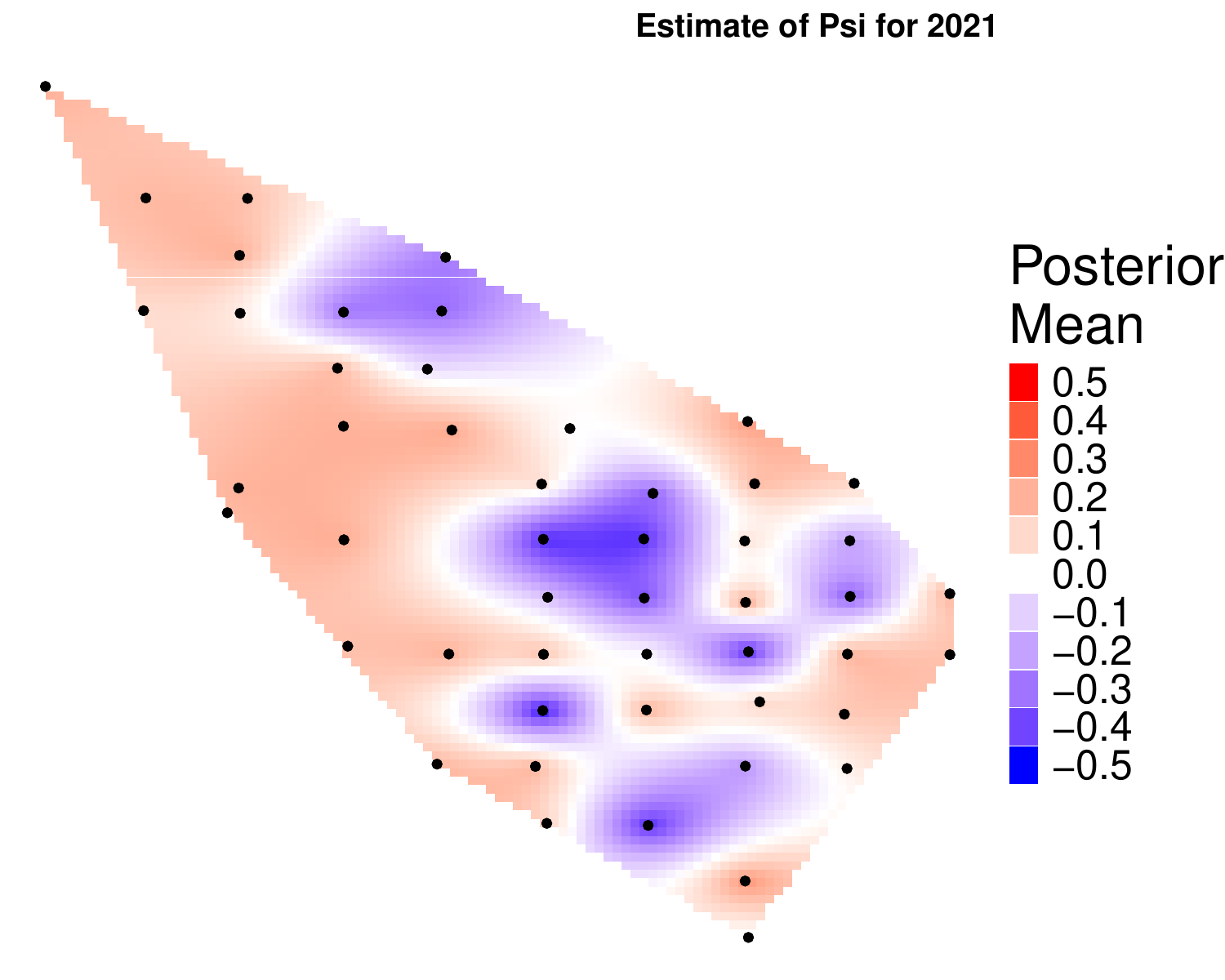}
        \includegraphics[width=0.32\linewidth]{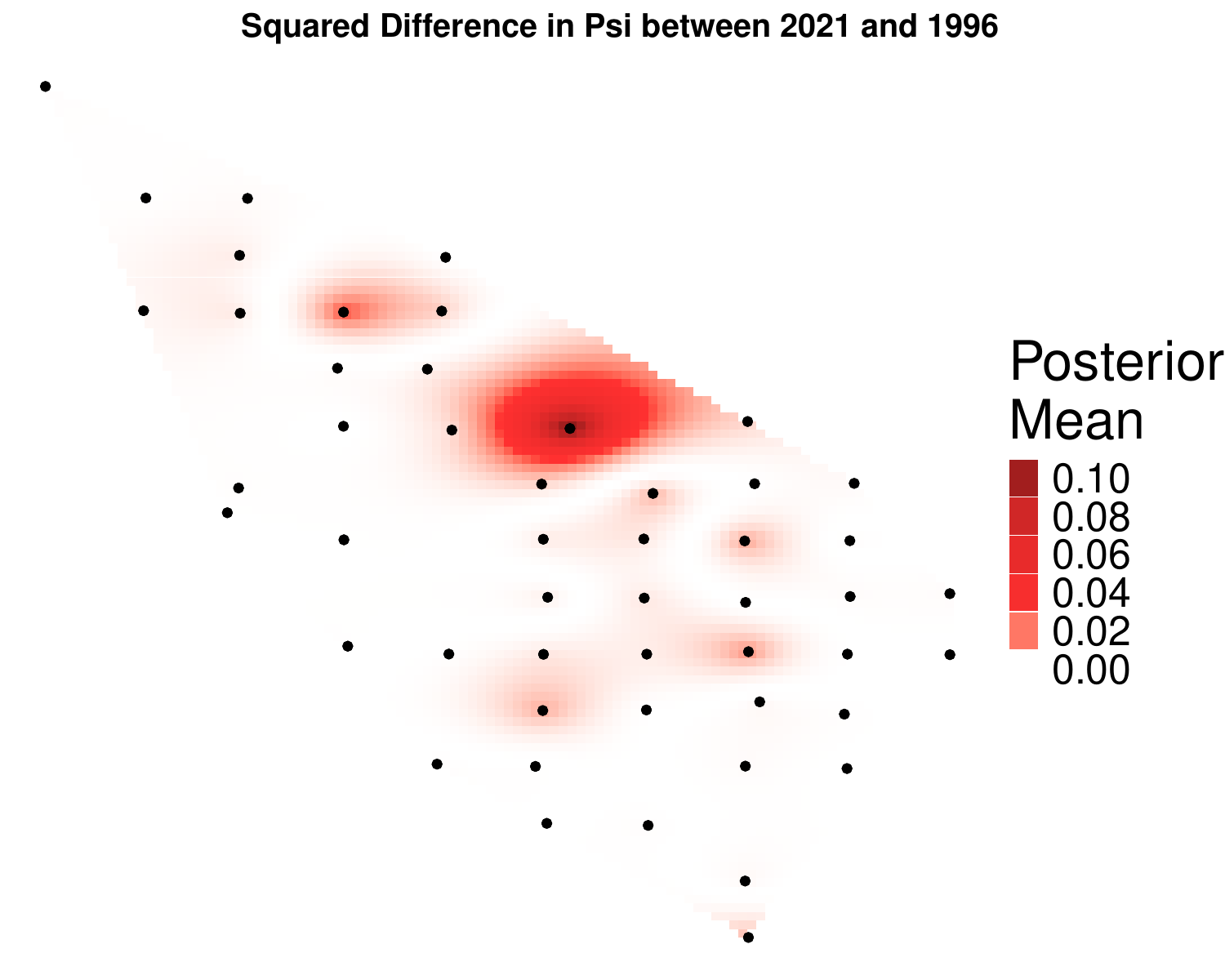}
\caption{Posterior mean surfaces for the spatial random effects, $\psi_t(\bs)$, in 1996 and 2021. The left and middle panels show the posterior mean spatial random effect surfaces for each survey year, and the right panel shows the posterior mean squared difference, $(\psi_{2021}(\bs) - \psi_{1996}(\bs))^2$. Black points denote observed sampling locations.}    \label{fig:posterior_psi}
\end{figure}

We plot the posterior mean surfaces for the spatial random effects $\psi_t(\bs)$ for both years, as well as the squared difference between years, in Figure \ref{fig:posterior_psi}.  These residual spatial surfaces persist but evolve. In both 1996 and 2021, the estimated $\psi_t(\bs)$ surfaces show localized regions of positive and negative spatial deviation, suggesting that there remains spatially structured variation in community composition not fully explained by the measured environmental covariates and geographic distance. The broad spatial patterns are similar across years. For 1996, plots with negative spatial random effects tended to have higher graminoid cover (likely more members of the Restionaceae), lower small shrub cover (likely more members of the Ericaceae) and lower species richness. The patterns for 2021 are similar but with lower small shrub cover was less salient.  These random effects suggests that our current model may not include factors relevant to factors that drive the distribution of members of the Restionaceae and Ericaceae family such as factors affecting seed bank persistence \citep{holmes_patterns_2004} and surface hydrological factors at finer resolution than our current moisture class scheme \citep{araya_fundamental_2011, silvertown_experimental_2012, lo_determination_2021}. The association of these plots with species richness is also interesting in that it suggests that differing mechanisms affect turnover and local diversity. 

The squared difference surface corresponds to the within years random effects in our model and reveals some regional changes between 1996 and 2021.  This indicates that the latent spatial structure is not static through time. These patterns support the use of a bivariate spatial process for $\psi_t(\bs)$, allowing the two years to be correlated while still permitting year-specific spatial deviations. The plots with high squared differences tended to have lower graminoid cover (~10\% lower) and no change in geophyte cover throughout both time periods. Given that many graminoids are germinated via smoke \citep{holmes_patterns_2004} and geophytes are sensitive to fire, this suggests that perhaps these plots had an unaccounted for factors such as larger rocky outcroppings serving as local fire refuges.

However, it is important to note that the space-time random effects contribution is through the $\eta_{t,t'}(\bs,\bs')$.  With only two time points, we have three random effects surfaces: $\eta_{1,1}(\bs, \bs')$, $\eta_{2,2}(\bs, \bs')$ and $\eta_{1,2}(\bs, \bs')$ since $\eta_{1,2}(\bs, \bs') = \eta_{2,1}(\bs', \bs)$.  We illustrate how the spatial random effects enter into model by considering a fixed reference site $\bs_0$ in 1996. This selection is illustrative and could be generated for any point in either year.

\begin{figure}[h]
    \centering
    \includegraphics[width=0.32\linewidth]{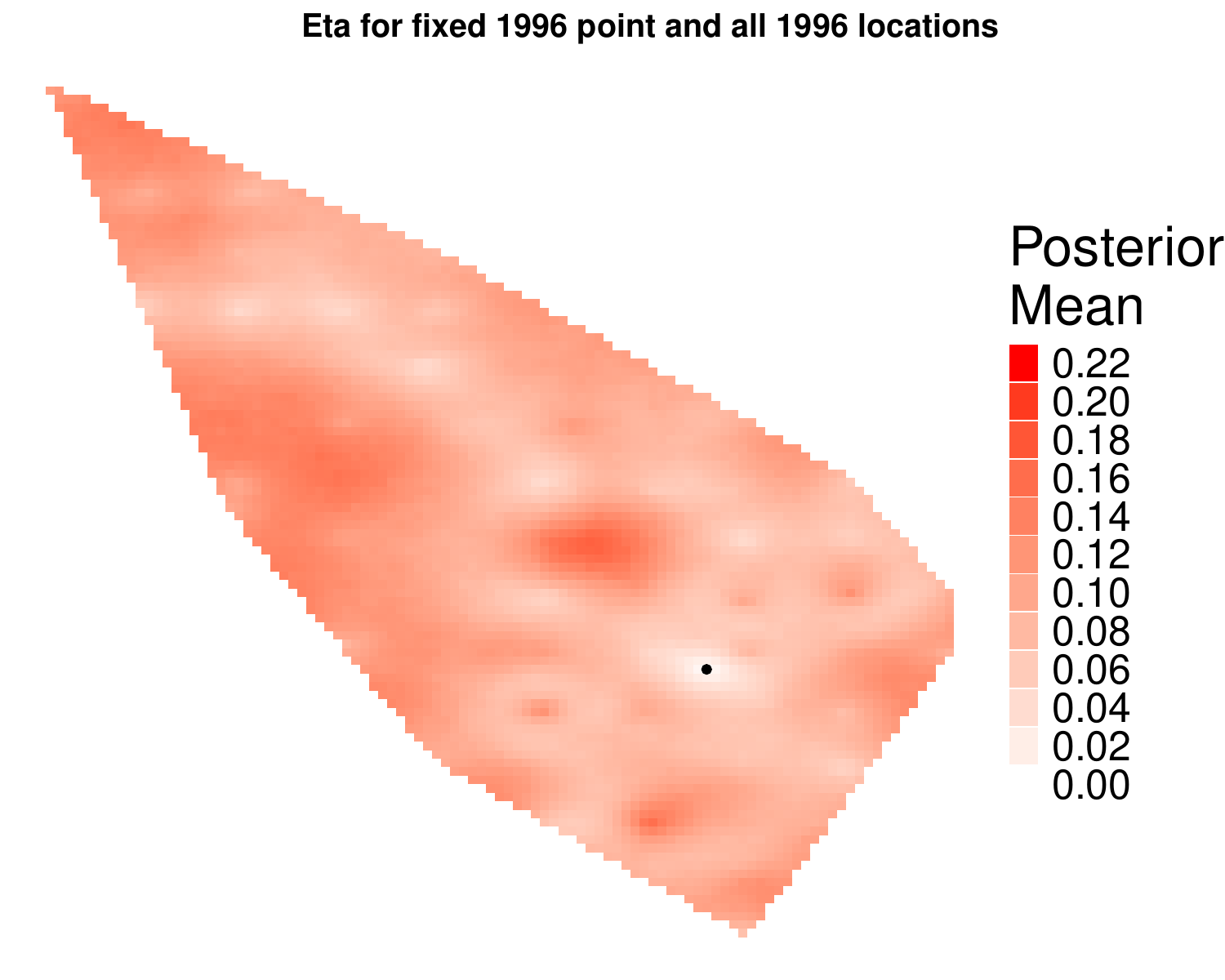}
    \includegraphics[width=0.32\linewidth]{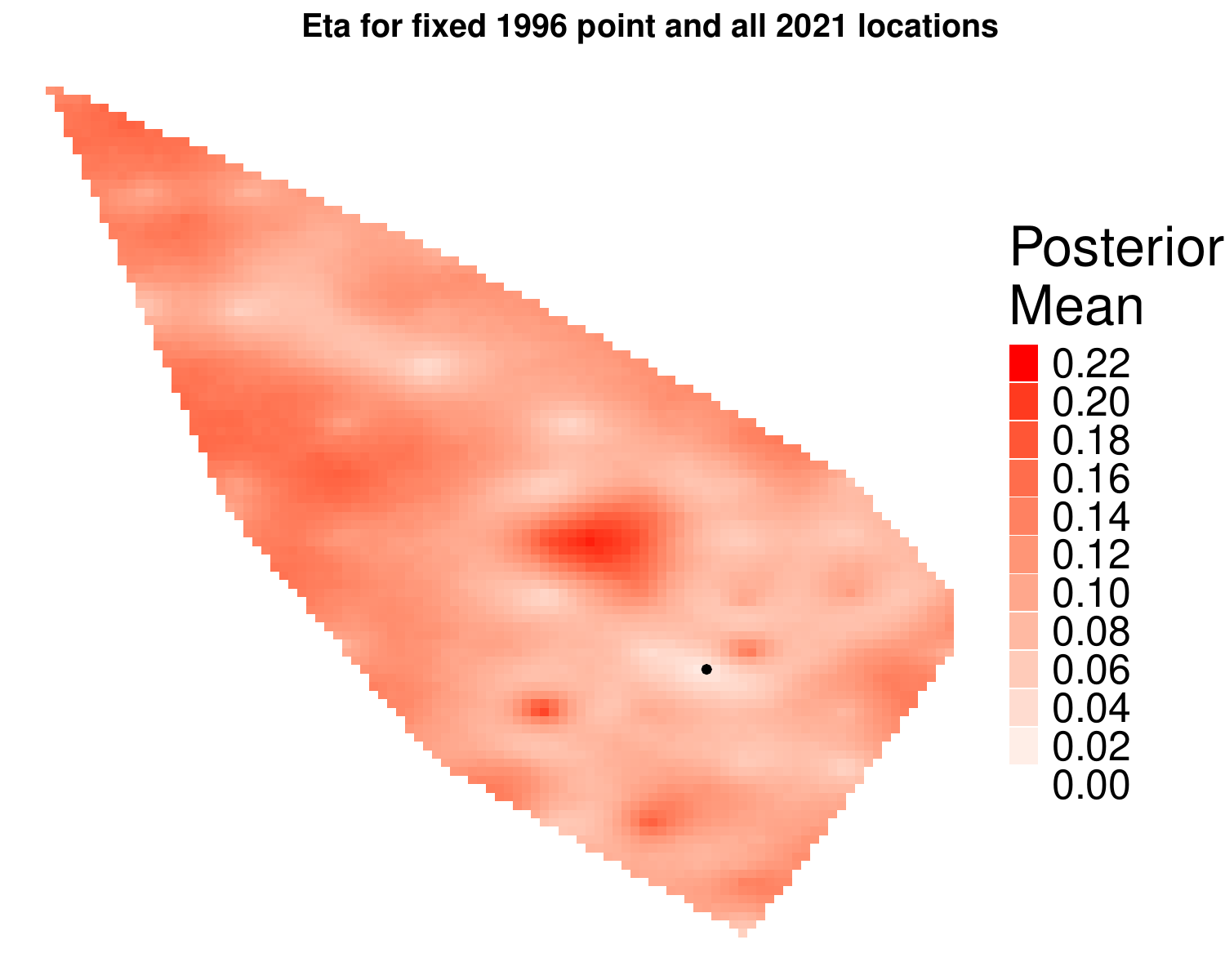}
        \includegraphics[width=0.32\linewidth]{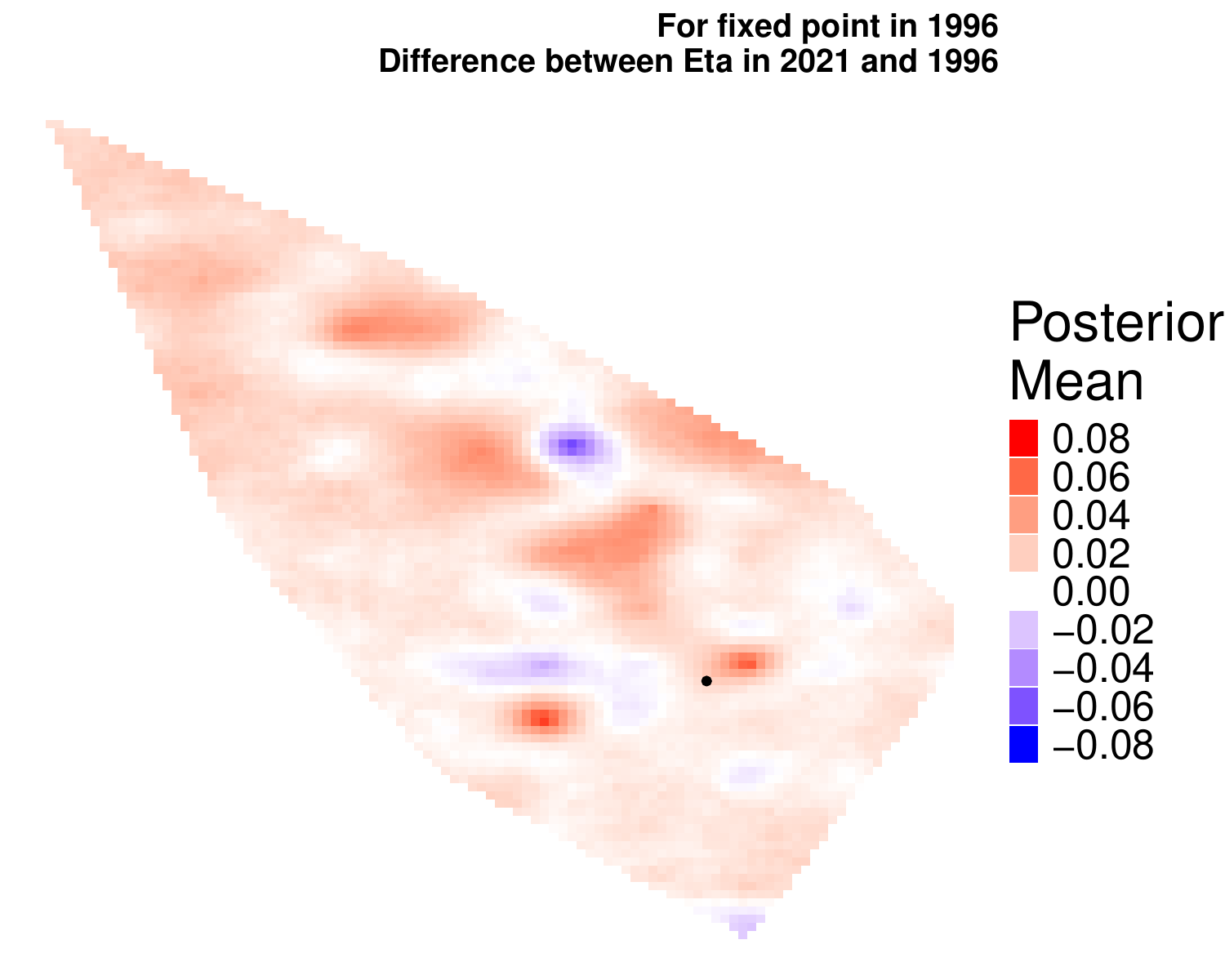}

\caption{Contribution of the space-time random effect to pairwise dissimilarity for a fixed reference site in 1996. The left panel shows $\eta_{1996,1996}(\bs_0,\bs)$ for the fixed reference site $\bs_0$ and all 1996 locations, while the middle panel shows $\eta_{1996,2021}(\bs_0,\bs)$ for the same reference site and all 2021 locations. The right panel shows the difference between the 2021 and 1996 $\eta$ surfaces.}    \label{fig:posterior_eta}
\end{figure}

The $\eta_{t,t'}(\bs,\bs')$ surfaces in Figure \ref{fig:posterior_eta} show how the spatial random effects contribute to pairwise dissimilarity model. Because $\eta_{t,t'}(\bs,\bs') = \{\psi_t(\bs) - \psi_{t'}(\bs')\}^2$, larger values occur where the latent spatial effect at the comparison location differs more strongly from the latent effect at the fixed reference site. For both the 1996 and 2021 comparison surfaces, $\eta$ is smallest near locations with similar latent spatial effects and larger in regions where the latent process differs from the reference site. These differences are spatially structured residual dissimilarities beyond that explained by the environmental covariates or distance between sites. The difference plot shows that the random-effect contribution to dissimilarity changes between years, but the magnitude of these changes is generally modest relative to the overall level of $\eta$. This suggests that the latent spatial contribution is temporally correlated but not identical across surveys, consistent with the separable bivariate Gaussian process specification for $\psi_t(\bs)$.


\section{Summary and Future Work}\label{sec:summary}

We studied biodiversity change through beta diversity, using Bray--Curtis dissimilarity to compare species composition across sites and years. Building on the spatial GDMM framework of \cite{white2024generative}, we developed a dynamic spatiotemporal extension that jointly models dissimilarities across both space and time. A primary advantage of this approach is that it used the full collection of pairwise site-year dissimilarities, including both within-year and between-year comparisons, rather than analyzing beta diversity separately by year. We proposed a generative model that accommodates boundary values corresponding to perfect similarity and perfect dissimilarity through a latent dissimilarity process, and we allowed the mean and variance structure to evolve through time-dependent warping functions, dynamic spatial random effects, and heterogeneous residual variability.

To accommodate this richer data structure, the proposed model introduces four key elements. First, a global between-year effect captured systematic differences between within-year and between-year dissimilarities. Second, evolving distance and environmental warping functions allowed covariate effects on beta diversity to change over time. Third, ordinal and categorical predictors were incorporated directly into the dissimilarity framework. Finally, space-time random effects accounted for residual spatial and temporal dependence not explained by measured environmental variables. Together, these components provided a unified framework for modeling spatial beta diversity, temporal beta diversity, and their interaction in a single generative model.

We applied our modeling to a collection of $50$ sites in the Cape Floristic Region of South Africa where site level species composition vectors were collected in 1996 and in 2021.  We were able to reveal the benefit of incorporating spatial separation, time dependent warping and inclusion of the random effects.

Conceptually, our modeling in Section \ref{sec:methods} can be extended to the setting of more than two time points, e.g., regularly sampled sites using autoregressive temporal specifications or irregularly sampled sites using continuous time specifications. See, e.g., \cite{banerjee2025hierarchical,wikle2019spatio} for details. 

With $m$ years and the same $n$ sites for each year, emulating the modeling in Section \ref{sec:methods}, we would need to introduce $m$ warping functions and variable importance parameters for each covariate. For the distance warping, we would need ${m + 1 \choose 2}$ warping functions and variable importance parameters to account for all combinations of years with replacement. The simple indicator function for the intercept would now need to accommodate $m$ years, either through dynamic models or Gaussian process models \citep[see][]{west1997bayesian, banerjee2025hierarchical}. Finally, we would need to specify as many as ${m + 1 \choose 2}$ variances.  We have a severe explosion of parameters and functions; practically, we would have to propose simplifying assumptions. We do not explore this challenge further here, but this provides several avenues for future research.



With regard to future work, several extensions appear promising. It may be useful to evaluate alternative dissimilarity measures, e.g., Raup-Crick or Hellinger distances, to determine whether the main ecological conclusions are sensitive to the choice of beta-diversity metric. For example, metrics may be dependent on temporal order, which would permit the model to distinguish, for example, between change from 1996 to 2021 and change from 2021 to 1996. Datasets with more regularly spaced sampling intervals would allow for models with explicit spatio-temporal terms compared to the two time point case study we explore here. Such extensions would provide a natural way to further explore how different aspects of community change are expressed in time and space.

\section*{Acknowledgments}\label{sec:acknow}

This work was supported by National Science Foundation (NSF) grant DEB-1046328, with a Research Experiences for Undergraduates Broadening Participation (REU-BP) Supplement to J. A. Silander, a National Aeronautics and Space Administration grant (80NSSC22K1383) to J. A. Silander,  and the South African Environmental Observation Network (SAEON), a research facility of the National Research Foundation of South Africa. Additionally, H.A. Frye was supported by the Townsend lab at the University of Wisconsin-Madison during the writing of this manuscript with funding from NSF grant (DBI-2021898) and NASA grant (80NSSC22K0905).

\section*{Data and Code Availability}\label{sec:data_avail}

Code used to fit the stGDMM described in this paper, along with the processed site-year covariate and species-abundance data needed to reproduce the model fitting and results, are included as supplementary material accompanying this submission. The raw vegetation survey data are curated by the South African Environmental Observation Network (SAEON) Fynbos Node and the University of Cape Town, and are available from the corresponding author upon request.

\bibliographystyle{apalike}
\bibliography{refs}

\end{document}